\documentclass[aps,prfluids,reprint,onecolumn]{revtex4-2}
\usepackage{graphicx}
\usepackage{dcolumn}
\usepackage{bm}

\usepackage{xcolor}
\usepackage{amsmath}
\usepackage{siunitx}
\usepackage[caption=false]{subfig}

\newcommand{\pd}[2][]{\partial_{#2} #1}

\newcommand{\integ}[4]{\int_{#1}^{#2}{\, #3 \, \mathrm{d} #4}}
\newcommand{\dinteg}[7]{\int_{#1}^{#2} \!\!\! \int_{#3}^{#4}{\, #5 \, \mathrm{d} #6 \,\mathrm{d} #7}}

\newcommand{\cauthor}[1]{\citeauthor{#1} \cite{#1}}

\newcommand{\ds}{\displaystyle}
\def\*#1{\bm{#1}}
\newcommand{\e}{\mathrm{e}}

\newcommand{\aphi}{\partial_\phi a}

\newcommand{\Nabla}{\boldsymbol{\nabla}}
\newcommand{\In}{\text{in}}
\newcommand{\Out}{\text{out}}
\newcommand{\ii}{\mathrm{i}}

\begin{document}

\title{Droplet motion on chemically heterogeneous substrates with mass transfer. II. Three-dimensional dynamics}
\author{Nikos Savva}
\email{Corresponding author: n.savva@cyi.ac.cy}
\affiliation{Computation-based Science and Technology Research Center, The Cyprus Institute, Nicosia 2121, Cyprus}
\author{Danny Groves}
\affiliation{School of Mathematics, Cardiff University, Cardiff, CF24 4AG, United Kingdom}

\begin{abstract}
We consider a thin droplet that spreads over a flat, horizontal and chemically heterogeneous surface. The droplet is subjected to changes in its volume though a prescribed,  arbitrary spatio-temporal function, which varies slowly and vanishes along the contact line. A matched asymptotics analysis is undertaken in the long-wave limit of the Stokes equations with slip to derive a set of evolution equations for the Fourier harmonics of nearly circular contact lines. Numerical experiments highlight the generally excellent agreement between the long-wave model and the derived equations, demonstrating that these are able to capture many of the features  which  characterize the intricate interplay between substrate heterogeneities and mass transfer on droplet motion.
\end{abstract}

\maketitle

\section{Introduction}\label{3DSec_Introduction}

When the size of a droplet changes as it spreads, a number of interesting effects emerge. Alongside the usual pinning, stick-slide and stick-jump modes which are observed in the absence of mass transfer, such effects also include, e.g.\ the constant-radius, constant-angle and the recently reported snapping modes (see, e.g.\ Refs.~\cite{Erbil2002,Lam2002,Dietrich2015,Stauber2015,Stauber2015a,Wells2018}; see also Ref.~\cite{Groves2021} and the references therein). Such effects are commonly attributed to local variations in the substrates' topographical and/or chemical features, which may temporarily trap the contact line \cite{Pradas2016,Groves2021}. Due to the highly nonlinear and multi-scale nature of the mechanisms underpinning the dynamics of contact lines, investigations based on purely computational approaches are highly non-trivial and are currently capable of only partly resolving the lengthscales present \cite{Afkhami2018,Lacis2020,Sui2014}.

Within the long-wave approximation and for weakly distorted contact lines, lower-dimensional models were derived for droplets of fixed volume that were capable of accurately capturing the evolution of contact lines in three dimensions~(3D), and at a fraction of the computing cost that would have been required for simulating the full long-wave model~\cite{Savva2018}. The present study offers an extension to Ref.~\cite{Savva2018} to perform, a matched asymptotic analysis for non-axisymmetric, 3D droplets of variable mass. It constitutes the second part of the work undertaken by the present authors which considers the same setting in two dimensions~(2D) (see Ref.~\cite{Groves2021}, hereinafter referred to as Part I). As in these aforementioned studies, we pursue a combined analytical and numerical investigation of a long-wave evolution equation for the droplet thickness, which is derived by considering thin viscous droplets in the gravity-free regime with strong surface tension effects and negligible inertia \cite{Greenspan1978}. Assuming that the spreading of the droplet and the rate of mass transfer is slow, analytical progress is possible with the method of matched asymptotic expansions in the limit of vanishingly small slip lengths. In this manner, the dynamics of the macroscale where capillary forces dominate are coupled with the processes governing the microscale where slip effects manifest themselves, to ultimately yield a set of simpler equations that approximates the full model \cite{Lacey1982,HOCKING1983}.

To avoid the complications of the analysis arising in the 2D counterpart where transcendental equations need to be solved for the contact point velocities \cite{Groves2021,OLIVER2015}, we assume that no fluid transfer occurs at the contact line so that explicit expressions can be derived for the contact line velocity. While this assumption is not appropriate for describing mass loss through evaporation since the evaporative flux is maximized there \cite{AJAEV2005,Saxton2016,Savva2017}, it is valid in other scenarios, e.g. when the liquid flux in the droplet is localized somewhere within its footprint. Besides, we have shown in Ref.\ \cite{Groves2021} that many of the interesting features observed in studies involving evaporating droplets are rather general, and characterize the dynamics captured by the present model. A key contribution of this work is that the more realistic 3D model allows us to make comparisons with experimental observations, so that features such as the stick-slip events observed by \citeauthor{Dietrich2015} for evaporating droplets \cite{Dietrich2015}, and the constant-radius and constant-angle modes observed by \citeauthor{Lam2002} for fluid pumping through a needle \cite{Lam2002}, can be explored by varying the parameters controlling mass transfer and chemical heterogeneity.

This study is structured as follows. In Sec.\ \ref{3DSec_ProblemFormulation} we present the model and its assumptions. Using the method of matched asymptotic expansions in Sec.\ \ref{3DSec_MatchedAsymptotics}, we derive a reduced model for the motion of the contact line that consists of a set of integro-differential equations, which approximates the dynamics of the full model of Sec.\ \ref{3DSec_ProblemFormulation}.  In Sec.\ \ref{3DSec_Results} the results of the analysis are scrutinized against numerical solutions to the governing partial differential equation (PDE) to extract insights on the interplay between mass transfer and surface heterogeneities, contrasting, where appropriate, the results in Part I of this study. Concluding remarks are offered in Sec.\ \ref{3DSec_Conclusion}.

\section{Problem Formulation}\label{3DSec_ProblemFormulation}

Consider a droplet in 3D moving on a flat, horizontal and chemically heterogeneous substrate under the action of capillary pressure and mass transfer effects. The problem formulation and assumptions closely follow Part I, so that the corresponding non-dimensional long-wave PDE which governs the evolution of the droplet thickness $h(\*x,t)$ is given by
\refstepcounter{equation}
$$\label{GovPDE}
	\pd[h]{t}+\Nabla\cdot\left[h(h^2+\lambda^2)\Nabla\nabla^2 h\right] = q, \eqno{(\theequation{\text{a}})}
$$
\begin{figure}[t!]
	\centering
	\includegraphics[scale = 1]{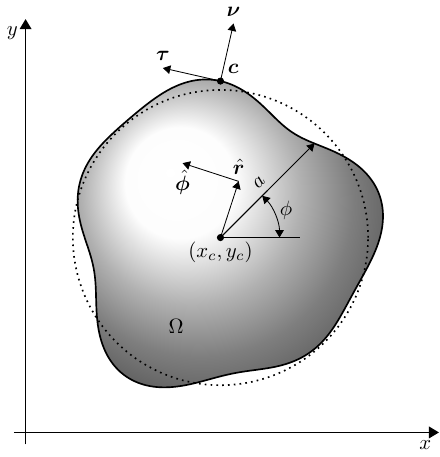}
	\caption{A top view of the geometry highlighting the wetted region $\Omega(t)$ (in shaded gray). The contact line $\*c$ is described by the polar curve $a(\phi,t)$ in the coordinate system with origin at $(x_c,y_c)$ and corresponds to a perturbation from a circle (dotted curve). The vectors $\hat{\*r}$ and $\hat{\*\phi}$ represent the unit radial and azimuthal vectors in the moving frame, respectively; $\*\nu$ and $\*\tau$ denote the outward unit normal and tangent vectors to the contact line $\*c$, respectively.}
	\label{3DFig_CoordTransform}
\end{figure}%
where $q(\*x,t)$ models the mass transfer, hereinafter referred to as the liquid flux, $\lambda$ is the slip length and the gradient and Laplacian operators are defined in 2D.  Equation~(\ref{GovPDE}a) has been made non-dimensional by scaling the lateral scales with $\ell = [V/(2\pi\alpha_s)]^{1/3}$, $h$ with $\alpha_s \ell$, $\lambda$ with $\alpha_s \ell/\sqrt{3}$, $t$ with $\bar\tau=3\mu\ell/(\sigma\alpha_s^3)$, and $q$ with $\rho\alpha_s \ell/\bar\tau$. In these scalings, $\sigma$ is the surface tension, $\rho$ is the liquid density, $\mu$ is the viscosity of the fluid, and $V$ and $\alpha_s$ are some reference values for the volume and contact angle, respectively. Consistently with Part I, we have employed the inverse linear slip model, which, unlike the classical Navier slip model, does not exhibit logarithmic singularities for $\nabla^2h$ at the contact line, a desirable feature in simulations of Eq.~(\ref{GovPDE}a). However, it turns out that, within the orders of terms retained in the asymptotic analysis, the results are identical for both models \cite{Savva2011,Groves2021}.

The substrate is decorated with chemical heterogeneities, which are described  with locally varying contact angles according to $\theta(\*x)$, as scaled with respect to $\alpha_s$. Along the contact line, $\*x=\*c$, we have 
$$
h=0,\quad \left|\Nabla h\right|=\theta \quad\text{and}\quad	\left(\pd[\*c]{t}-\lambda^2\Nabla\nabla^2 h\right)\cdot\*\nu =q/\theta,
\eqno{(\theequation\text{b,c,d})}\label{BCs}
$$
where $\*\nu$ denotes the outward unit normal vector to the contact line, which is in the plane of the substrate (see Fig.~\ref{3DFig_CoordTransform}).  We note here that, just as in the 2D case~\cite{Groves2021}, condition~(\ref{BCs}d) is of kinematic type and arises from a local expansion of Eq.~(\ref{GovPDE}a) near $\*x=\*c$ so that Eqs.~(\ref{BCs}b) and (\ref{BCs}c) together with
$$
\frac{\mathrm{d}}{\mathrm{d}t}\integ{\Omega(t)}{}{h(\*x,t)}{\*x} = \integ{\Omega(t)}{}{q(\*x,t)}{\*x} = \dot{v}(t)
\eqno{(\theequation\text{e})}
$$
simultaneously hold (see Appendix~\ref{sec:BC} for an outline of the derivation of Eq.~(\ref{BCs}d)). In Eq.~(\ref{BCs}e), $\Omega(t)$ and $v(t)$ denote, respectively, the wetted region of the substrate and the (dimensionless) droplet volume, and the dot denotes differentiation with respect to $t$.  In the present study we assume that $q\to0$ as $\*x\to\*c$, so that Eq.~(\ref{BCs}d) reduces to $\left(\pd[\*c]{t}-\lambda^2\Nabla\nabla^2 h\right)\cdot\*\nu=0$ along the contact line. In this manner, as mentioned in Sec.~\ref{3DSec_Introduction}, we avoid having to work with transcendental equations like the ones which arise if such analysis is carried out in 2D~\cite{OLIVER2015,Groves2021}. In this context, having vanishingly small fluxes along the contact line is realistic if, say, the mass flux is localized somewhere within the droplet. Thus, solving the long-wave model~\eqref{GovPDE} determines how the droplet thickness, $h$, and contact line position, $\*c$, evolve in time for prescribed contact angles $\theta$, mass transfer $q$, and slip length $\lambda$. In this work, $q$ may be taken to be an arbitrary spatiotemporal function, whose contributions are dominant within the droplet footprint and vanish along the contact line. This situation can model, for example, localized fluxes, e.g. by pumping fluid through a needle. For the analysis that follows, we introduce the normalized flux $\tilde{q}$, defined as $q = \dot{v} \tilde{q}$, which, in accordance with Eq.~(\ref{GovPDE}e), must satisfy
\begin{equation}
\integ{\Omega(t)}{}{\tilde{q}(\*x,t)}{\*x} =1. \label{eq:qtilde}
\end{equation}
In this manner, we may describe processes where we can control how the volume of the droplet changes in time (e.g. by artificially pumping fluid in and out of the droplet), as opposed to, say, evaporative effects, where volume variations are not explicitly controllable, but emerge from natural processes. To facilitate the analysis, the free-boundary problem~\eqref{GovPDE} is transformed to a problem defined on a disk of unit radius, centered at $\*x=\*x_c(t)$ \cite{Savva2018}. In principle, the choice for $\*x_c(t)$ can be arbitrary as long as it is contained within $\Omega(t)$, so that we are able to define the following mapping
\begin{equation}\label{3D_Transformation}
\*x=\*x_c(t) + r a(\phi,t) \hat{\*r},
\end{equation}
where $0\le r\le1$, $0\le\phi<2\pi$, $a(\phi,t)$ is assumed to be a single-valued function of $\phi$ and $\hat{\*r}=(\cos\phi,\sin\phi)$. At the contact line, for which $r=1$, we have that 
\begin{equation}\label{3D_CL}
\*c=\*x_c+a\hat{\*r}.
\end{equation}
With the transformation \eqref{3D_Transformation}, the original problem \eqref{GovPDE} reduces to one of determining $h(r,\phi,t)$, $a(\phi,t)$ and $\*x_c(t)$ by casting their corresponding PDEs and conditions in the new $(r,\phi)$-coordinate system. Lastly, we assume that  $\*x_c(t)$ evolves such that the origin of the moving frame is always located at the centroid of the wetted area $\Omega(t)$,  which requires that
\begin{equation}\label{3D_CentroidConditions}
\int_0^{2\pi}a^3\cos\phi\,\mathrm{d}\phi=0,\quad\int_0^{2\pi}a^3\sin\phi\,\mathrm{d}\phi=0.
\end{equation}
From Eqs.~\eqref{3D_CentroidConditions} we are able to obtain an evolution of equation for each of the two components of $\*x_c(t)$  (see Ref.~\cite{Savva2018} for more details on the transformation of the temporal and spatial derivatives in the new coordinate system, and the resulting equations for $\dot{x}_c(t)$ and $\dot{y}_c(t)$). As in Part I, we will pursue a combined analytical--numerical approach to deduce lower dimensional models for the evolution of the contact line, given that as $\lambda$ assumes realistically small values, the curvature of the free surface shape steepens near the contact line. Hence, the full problem~\eqref{GovPDE} becomes stiff due to these boundary layers which require denser computational meshes to be resolved, therefore making simulations more demanding~\cite{Savva2018}.

\section{Matched Asymptotic Analysis}\label{3DSec_MatchedAsymptotics}

Just as in the case for droplets of constant volume~\cite{Savva2018}, we
assume that $a\gg\partial_{\phi}a$, expanding $a(\phi,t)$ as a truncated
Fourier series of the form
\begin{equation}
\label{3D_CLExpansion}a(\phi,t) = \sum_{m=0}^{M} a_{m}(t)\mathrm{e}%
^{\mathrm{i} m\phi},
\end{equation}
where $a_{m}$ are generally complex functions of $t$ to be determined, with $|a_{m}(t)|\ll a_{0}(t)$, so that $a(\phi,t)$ describes weak perturbations from a circular contact line of radius $a_{0}(t)$. The series is truncated at $M\ll\lambda^{-1}$ in order to suppress non-physical behaviors that may arise in the asymptotic model if we allow for contact line variations  occurring at lengthscales that are smaller than $\lambda$~\cite{Savva2018}. We note that the complex series representation in Eq.~\eqref{3D_CLExpansion} and all expressions containing complex exponentials that are going to be discussed hereinafter are considered with their imaginary parts discarded.

The same assumptions put forth in the analysis of Part I of this study hold, namely that we focus on slow droplet dynamics occurring on the slow time scale $\tau=\epsilon t$, where $\epsilon=1/\ln\lambda^{-1}\ll1$ as $\lambda\rightarrow0$, assuming that both $|\partial_{t}\*c|$ and $\dot{v}$ are $O(\epsilon)$. Provided that there is sufficient separation of scales (e.g., for droplets away from equilibrium, and when $v(t)\gg\lambda$ and $\lambda|\partial_{\phi}a|\ll1$), we may develop a lower-dimensional model based on matched asymptotic expansions that adequately describes the dynamics for all $t$, without having to construct some composite expansion to capture all the pertinent time scales. Before embarking on such analysis, it is important to note that, unlike the numerics, $\*x_{c}$ is chosen differently for the analysis. In particular, we require that for any given contact line shape $\*c$, the coordinates $\*x_c(t)=(x_c(t),y_c(t))$ are chosen such that the first harmonic of $a(\phi,t)$ is always suppressed so that $a_{1}(t)\equiv0$, associating  $\e^{\ii \phi}$  terms with the origin of the moving frame. Uniquely defining $\*x_{c}$ in this manner is always possible for all single-valued functions $a(\phi,t)$. Although for such a choice for the coordinate system the contact line shape is no longer centered at the centroid of the drop, Eq.~\eqref{3D_CLExpansion} in Eqs.~\eqref{3D_CentroidConditions} gives $\text{Re}(a_1)+O(a_m^2) = \text{Im}(a_1)+O(a_m^2)=0$, readily implying that the two choices are equivalent to each other, at least within the orders of terms retained in the asymptotics.

For the analysis, we introduce the smallness parameter $\delta\ll1$ as a characteristic size of the azimuthal disturbances of the contact line. In this setting,  $\dot{a}_0$, $\dot{x}_c$ and $\dot{y}_c$ are all assumed to be $O(\epsilon)$ quantities, whereas $\dot{a}_k$, $k>1$ are $O(\delta\epsilon)$.  Thus, by treating $\epsilon$ and $\delta$ as ordering parameters in the asymptotics to identify the various terms that need to be retained,  we introduce the $O(1)$ variables
\begin{equation}
b_k(t) =\left\{
\begin{array}
[c]{ll}%
a_{0}\left(  t\right)   & \text{for }k=0\\
0, &\text{for } k=1\\
a_{k}\left(  t\right)  /\delta & \text{for }k\ge2
\end{array}
\right.  ,\quad  u_k(t)=\frac{1}{\epsilon}\times\left\{ 
\begin{array}[c]{ll}
\dot{b}_k,&\text{for } k\neq1\\
\dot{x}_{c}\left(  t\right)  -\mathrm{i}\dot{y}_{c}\left(  t\right),&\text{for }{k=1}  
\end{array}
\right..
\end{equation}
As in Ref.~\cite{Savva2018}, the analysis pursued entails deducing evolution equations for $\*x_{c}$ and the Fourier coefficients of $a(\phi,t)$ to approximate the corresponding evolution of $\*c$ of the full problem~\eqref{GovPDE} as $\lambda\rightarrow0$. This is done by probing separately into the dynamics at the microscale, the \textit{inner region}, and the macroscale, the \textit{outer region}, developing the  asymptotic expansions of $\partial_{\nu}h=\*\nu\cdot\boldsymbol{\nabla} h$ corresponding to each region as the other is approached. Obtaining the evolution equations for $\*c$ is achieved by matching the two expansions in the distinguished limit $\lambda\ll\delta^2\ll\epsilon\ll\delta\ll1$. Although certain parts of the analysis that were carried out in the aforementioned work of Savva \emph{et al.} are applicable in this setting as well, here we include the contributions of the mass flux term, $q$ in  Eq.~\eqref{GovPDE} and additional correction terms coming from the axisymmetric mode and the centroid motion that were previously omitted.

\subsection{Inner region}

The dynamics of the inner region may be examined by sufficiently zooming into the vicinity of the contact line through the appropriate stretching transformations. Given that, by assumption, $q(\*c,t)=0$, mass transfer effects contribute to the $O(\lambda)$ terms, which are neglected here. Hence, the analysis is identical to the inner region asymptotics for droplets of constant volume \cite{Savva2018}. The key idea behind the inner region asymptotics in 3D is that if $|\partial_t\*c\cdot\*\nu|\gg\lambda$ and the contact line varies at lengthscales that are longer than slip (equivalently, $\lambda|\partial_\phi a|\ll1$ and $\lambda|\partial_\phi^2 a|\ll1$), the analysis of the inner region is the same as in the 2D geometry at leading order as $\lambda\to0$. The 2D geometry has been tackled in Ref.~\cite{Vellingiri2011}, but for the case of the Navier slip. As previously mentioned (see also Part I) both models possess the same leading-order asymptotics \cite{Savva2011,Sibley2014}, giving rise to the following two-term expansion for the inner region, $h_\In$,
\begin{equation}\label{eq:inner}
	-\pd[h_\In]{\nu} \sim \theta_*+\epsilon\frac{\partial_\tau{\*c}\cdot\*\nu}{\theta_*^2}\ln\left[\frac{\e\theta_*(\*c-\*x)\cdot\*\nu}{\lambda}\right]+\cdots \quad\text{as}\quad\frac{(\*c-\*x)\cdot\*\nu}{\lambda}\to\infty,
\end{equation}
where dots denote the higher-order corrections which are omitted and $\theta_* = \theta|_{\*x=\*c}$ is the  heterogeneity evaluated along the contact line. Combining Eqs.\ \eqref{3D_CL} and~\eqref{3D_CLExpansion}, and noting that $\*\nu=\hat{\*r}-\partial_\phi a\, \hat{\*\phi}/a_0+\cdots$ where $\hat{\*\phi} = (-\sin\phi, \cos\phi)$, we can produce an expansion for $\partial_\tau\*c\cdot\*\nu$ in the limit of nearly circular contact lines as
\begin{equation}\label{3D_CdotExpans}
	\partial_\tau{\*c}\cdot\*\nu = u_0+\left(u_1-\delta\frac{b_2}{b_0}u_1^*\right)\e^{\ii \phi}+\delta\sum_{m=2}^M\left[u_m+\frac{(m-1)b_{m-1}}{2b_0}u_1-\frac{(m+1)b_{m+1}}{2b_0}u_1^*\right]\e^{\ii m\phi}+\ldots
\end{equation}
where $b_{M+1}=0$ and superscripted stars denote complex conjugation. Using Eq.~\eqref{3D_CdotExpans}, we can write Eq.~\eqref{eq:inner} in terms of the moving polar coordinate system, Eq.~\eqref{3D_Transformation}, which is to be matched with the corresponding outer-region expansion. In the end, the expansion in Eq.~\eqref{eq:inner} becomes
\begin{equation}\label{3D_InnerSlopeFinal}
-\pd[h_\In]{\nu}\!\sim\!\theta_*\!+\!\frac{\epsilon}{\theta_*^2}\ln\left[\e a\theta_* \frac{1-r}{\lambda}\right]\left\{u_0+\left(u_1-\delta\frac{b_2}{b_0}u_1^*\right)\e^{\ii\phi}\!+\!\delta\sum_{m=2}^M\left[u_m\!+\!\frac{(m-1)b_{m-1}}{2b_0}u_1\!-\!\frac{(m+1)b_{m+1}}{2b_0}u_1^*\right]\e^{\ii m\phi}\right\}
\end{equation}
as $(1-r)/\lambda\to\infty$.

\subsection{Outer region}\label{sec:outer}

Slip effects are neglected in the outer region, thus dropping $O(\lambda)$ terms. Introducing the change of variable $\tau=\epsilon t$  transforms Eq.~\eqref{GovPDE} to
\begin{subequations}
\begin{equation}
\boldsymbol{\nabla}\cdot\left[  h_{\Out}^{3}\boldsymbol{\nabla}\nabla
^{2}h_{\Out}\right]  =\epsilon\left[  v^{\prime}\tilde{q}+\left(
\boldsymbol{x}_{c}^{\prime}+r\partial_{\tau}a\boldsymbol{\hat{r}}\right)
\cdot\boldsymbol{\nabla}h_{\Out}-\partial_{\tau}h_{\Out}\right]  ,
\end{equation}
where primes denote differentiation with respect to $\tau$, and this is solved subject to the conditions
\begin{align}
h_{\text{out}}(1,\phi,t) &  =0,\\
\int_{0}^{2\pi}\!\!\!\int_{0}^{1}%
{\,ra^{2}h_{\text{out}}\ \,\mathrm{d}r\,\mathrm{d}\phi} &  =v(\tau).
\end{align}
\end{subequations}
We then consider a quasistatic expansion of the form
\begin{equation}
h_{\text{out}}(r,\phi,\tau)=h_{0}(r,\phi,\tau)+\epsilon h_{1}(r,\phi
,\tau)+\ldots,\label{eq:h_expansion}
\end{equation}
where $h_{1}$ is assumed to be linear in $\delta$. The first term, $h_{0}$, describes the leading-order shape of the droplet in the bulk, and is determined by solving the quasiequilibrium problem
\begin{subequations}
\label{3D_LeadingOrderSystem}%
\begin{align}
\nabla^{2}h_{0}(r,\phi,\tau) &  =k(\tau),\\
h_{0}(1,\phi,\tau) &  =0,\label{3D_LeadingOrderCond1}\\
\int_{0}^{2\pi}\!\!\!\int_{0}^{1}{\,ra^{2}h_{0}\,\mathrm{d}r\,\mathrm{d}\phi}
&  =v(\tau).\label{3D_LeadingOrderCond2}%
\end{align}
\end{subequations}
We can straightforwardly obtain $h_{0}$ using a perturbation series in $\delta$, deducing $k\left(  \tau\right)  $ from the volume constraint, Eq.~\eqref{3D_LeadingOrderCond2}. This yields
\begin{equation}\label{3D_LeadingOrderOuter0}
	h_0(r,\phi,\tau) = \bar{\vartheta}\left[\frac{b_0(1-r^2)}{2}+\sum_{m=2}^{M}b_m\left(r^m-r^2\right)\e^{\ii m\phi}\right]+\ldots,
\end{equation}
where $\bar{\vartheta}(t)=4v(t)/(\pi b_0^3)$ denotes the (time-dependent) average value of the apparent contact angle $\vartheta$, which may be determined from the normal derivative of $h_0$ and is expanded as
\begin{equation}\label{3D_ApparentContactAngle}
	\vartheta = -\pd[h_0]{\nu}|_{\*c} = \bar{\vartheta}\left[1+\sum_{m=2}^M\frac{b_m(1-m)}{b_0}\e^{\ii m\phi}\right]+\ldots.
\end{equation}
In the analysis that follows, the expression for $h_0$ in Eq.~\eqref{3D_LeadingOrderOuter0} is rearranged to factor out $a$ and $\vartheta$, so that 
\begin{equation}
h_{0}\left(  r,\phi,\tau\right)  =\vartheta a\left[  h_{0,0}\left(  r\right)
+\delta\sum_{m=2}^{M}\frac{b_{m}  }{b_{0}  }h_{0,m}\left(  r\right)  \mathrm{e}^{\mathrm{i}m\phi}%
+\cdots\right]  ,\label{3D_LeadingOrderOuter}%
\end{equation}
where
\begin{align}
h_{0,0} (r) & =\frac{1}{2}\left(  1-r^{2}\right),  \\
h_{0,m} (r) & =r^{m}-1+\frac{m}{2}\left(  1-r^{2}\right).
\end{align}
The next order term in \eqref{eq:h_expansion}, $h_1$, captures the contributions due to mass transfer as well as the motion of the contact line. It satisfies the following linear partial differential equation
\begin{equation*}\refstepcounter{equation}\label{Eq_h1PDE}
	\Nabla\cdot \left(h_0^3\Nabla\nabla^2 h_1\right) = v'\tilde{q}-\pd[h_0]{\tau}+(\*x'_c+r\pd[a]{\tau}\hat{\*r})\cdot\Nabla h_0, \eqno{(\theequation{\text{a}})}
\end{equation*}
with conditions
\begin{equation*}
	h_1(1,\phi,t) =0,\quad \dinteg{0}{2\pi}{0}{1}{ra^2h_1}{r}{\phi} = 0.\eqno{(\theequation{\text{b,c}})}
\end{equation*}
The analysis of $h_1$ revolves around the truncated Fourier series for $h_1(r,\phi,\tau)$ and $\tilde{q}$, which are written as
\begin{align}
h_1(r,\phi,\tau) &= \frac{a}{\vartheta^2}\left(h_{1,0}(r,\tau)+h_{1,1}(r,\tau)\e^{\ii\phi}+\delta\sum_{m=1}^M \tilde{h}_{1,m}(r,\tau)\e^{\ii m\phi}\right),\label{eq:h1q}\\
\tilde{q}(r,\phi,t) &=\vartheta \left(q_0(r,t) + q_1(r,t)\e^{\ii\phi} + \delta \sum_{m=2}^M q_m(r,t)\e^{\ii m\phi}\right),\label{eq:qm}
\end{align}
where $q_m(1,t)=0$, $m\ge0$, since, by assumption, $q\to0$ as $\*x\to\*c$. The specific form of the series of $h_1$ captures the  dominant axisymmetric and centroid motions, the $h_{1,0}$ and $h_{1,1}$ terms, respectively, as well as the higher-order azimuthal terms, $\tilde{h}_{1,m}$. It is important to note that in Eqs.~\eqref{3D_LeadingOrderOuter}, \eqref{eq:h1q} and \eqref{eq:qm}   we have factored out expressions involving $a$ and $\vartheta$, so that the analysis that follows can be more concisely presented. The goal is to construct the asymptotic expansion of the normal derivative of $h_\Out$ as $r\to1^-$, cast in the form
\begin{equation}\label{eq:houtas}
-\partial_\nu h_{\Out} \sim \vartheta - \frac{\epsilon}{\vartheta^2} \left(\partial_r h_{1,0}(r,\tau)+\partial_r h_{1,1}(r,\tau)\e^{\ii\phi}+\delta\sum_{m=1}^M \partial_r \tilde{h}_{1,m}(r,\tau)\e^{\ii m\phi}\right).
\end{equation}
To achieve this, a series of laborious algebraic manipulations are required, some details of which are relegated to the Appendices that follow the main text. Firstly, we obtain in Appendix~\ref{sec:BVPs} a set of linear boundary value problems for  $h_{1,0}$, $h_{1,1}$ and $\tilde h_{1,m}$, $m\ge1$. In Appendix~\ref{sec:asympt}, we examine their derivatives with respect to $r$ as $r\to1^-$ in accordance with Eq.~\eqref{eq:houtas}. Consistently with relevant works in the literature, we find that the expansion in  Eq.~\eqref{eq:houtas} consists of logarithmically diverging terms as $r\to1^-$ and $r$-independent terms, which may be cast in the form of integrals, see Appendix~\ref{sec:asympt}. For further insights  into their form, a discussion on their large-$m$ asymptotics is offered in  Appendix~\ref{sec:integral}.

Anticipating ahead the matching which is to be performed for the cubes of the normal derivatives, we write the asymptotics of Eq.~\eqref{eq:houtas} as $r\to1^-$ in the form 
\begin{align}
-&\left(\partial_\nu h_{\Out}\right)^3  \sim \vartheta^3 +3 \epsilon\left\{\partial_\tau \*c\cdot\*\nu \ln (1-r) +\beta_0u_0+\beta_1u_1\e^{\ii\phi}+\delta\sum_{m=1}^M \left[\chi_m\beta_m u_m -  \frac{m\tilde{\beta}^0_m}{b_0}b_mu_0 +\frac{m-1}{2b_0}(\gamma_m-\tilde\beta_m^-)b_{m-1}u_1\right.\right.\nonumber\\
&\left.\left.\left.-\frac{m+1}{2b_0}(2\beta_m-\gamma_m+\tilde\beta_m^+)b_{m+1}u_1^* + v'\left(\zeta_m - \frac{3}{2v} \left(2m\zeta_0 b_m +\zeta_1 (m-1) b_{m-1} + (m+1) \zeta_1^* b_{m+1}\right)\right)\vphantom{\frac{m\tilde{\beta}^0_m}{b_0}}\right]\e^{\ii m\phi}\right.\vphantom{\sum_{m=1}^M}\right\},\label{3D_OuterSlopeFinal}
\end{align}
as derived from Eqs.~\eqref{eq:h11as}, \eqref{eq:h1mas} and \eqref{eq:h10as} together with Eqs.~\eqref{3D_CdotExpans} and \eqref{eq:tildebeta}. As discussed in Appendix~\ref{sec:integral}, the  $O(v'\zeta_0b_m)$, $O(v'\zeta_1^*b_{m+1})$ and $O(v'\zeta_1b_{m-1})$ terms in Eq.~\eqref{3D_OuterSlopeFinal} come from the large-$m$ asymptotics of the associated integrals. We opted against the much more involved route of capturing these terms by solving at each time the corresponding boundary value problems and extracting their pertinent asymptotics, which is deemed to be a reasonable compromise between accuracy and efficiency. In fact, in all the numerical tests we have performed, the omission of such terms has had very little impact on how well the derived asymptotic model agrees with full simulations of Eq.~\eqref{GovPDE}.

\subsection{Matching}
As in the 2D analysis conducted in Part I with no mass transfer at the contact points, we may asymptotically match the cubes of the inner and outer normal derivatives given by Eqs.~\eqref{3D_InnerSlopeFinal} and \eqref{3D_OuterSlopeFinal}. In this manner, and within the appropriate overlap region,  we have that $(\pd[h_\In]{\nu})^3\sim(\pd[h_\Out]{\nu})^3$ \cite{Sibley2015}. This allows us to eliminate the $\ln(1-r)$ terms, deducing a Cox--Voinov-type evolution  equation for the contact line from the remaining terms. Reverting to derivatives with respect to $t$ and the original variables for the harmonics, the equation obtained from matching takes the form
\begin{align}
\frac{\vartheta^3-\theta_*^3}{3}&=\partial_t\*c\cdot\*\nu\ln\left(\frac{\e a\theta_*}{\lambda}\right) -\sum_{m=0}^M \left[\beta_m U_m - m\tilde{\beta}_m^0 \frac{a_m}{a_0} U_0 +\frac{m-1}{2a_0}(\gamma_m-\tilde\beta_m^-)a_{m-1}U_1\right.\nonumber\\
&\left.\!-\frac{m+1}{2a_0}(2\beta_m-\gamma_m+\tilde\beta_m^+)a_{m+1}U_1^* + \dot{v}\left(\zeta_m \!-\! \frac{3}{2v} \left(2m\zeta_0 a_m +\zeta_1 (m-1) a_{m-1} +  \zeta_1^*(m+1) a_{m+1}\right)\right)\right]\e^{\ii m\phi},\label{eq:CLlaw}
\end{align}
where we set $U_m=\dot{a}_m$ for $m\neq1$ and $U_1=\dot{x}_c-\ii\dot{y}_c$.  From the Fourier coefficients of  Eq.~\eqref{eq:CLlaw}, we obtain differential equations for each of the components of $a(\phi,t)$ and the centroid dynamics, $\*x_c(t)$.  Specifically, we determine $U_0$ and $U_1$ from
\begin{subequations}\label{3D_ODE}
\begin{align}
(\psi_0+\hat\beta_0^0) U_0 &= w_0 + \dot{v}\zeta_0 \\
(\psi_0+\hat\beta_1^0)U_1 - \left[\frac{a_2}{a_0}\left( \psi_0 +\hat\beta_1^+ \right)-\psi_2\right] U_1^* &=w_1-U_0\psi_1 +\dot{v}\zeta_1
\end{align} 
whereas $U_m$ with $m>1$ satisfy
\begin{align}
(\psi_0+\hat\beta_m^0) U_m &= w_m  +\frac{1}{2}\left[ (m+1)\left(\psi_0+\hat\beta_m^+\right)\frac{a_{m+1}}{a_0}-\psi_{m+1} \right]U_1^* -\frac{1}{2}\left[ (m-1)\left(\psi_0+\hat\beta_m^-\right)\frac{a_{m-1}}{a_0}+\psi_{m-1} \right]U_1\nonumber\\&-\left( m\tilde{\beta}_m^0 \frac{a_m}{a_0}+\psi_m \right)U_0+\dot{v}\left[\zeta_m \!-\! \frac{3}{2v} \left(2m\zeta_0 a_m +\zeta_1 (m-1) a_{m-1} + (m+1) \zeta_1^* a_{m+1}\right)\right].
\end{align}
\end{subequations}
In Eqs.~\eqref{3D_ODE} we defined the constants
\begin{equation}
\hat\beta_m^0 =  1 -\ln \lambda- \beta_m,\quad \hat\beta_m^+ = 1 -\ln \lambda-2\beta_m+\gamma_m-\tilde\beta_m^+,\quad\hat\beta_m^-=1-\ln\lambda-\gamma_m+\tilde\beta_m^-,
\end{equation}
and the $\psi_m$ and $w_m$ correspond to the coefficients of the truncated Fourier series of $\ln(a\theta_*)$ and $(\vartheta^3-\theta_*^3)/3$, respectively,
\begin{equation}
\frac{\vartheta^3-\theta_*^3}{3} = \sum_{m=0}^M w_m\e^{\ii m\phi},\quad\ln (a\theta_*) = \sum_{m=0}^M \psi_m\e^{\ii m\phi}.
\end{equation}
Likewise, we set $\delta=1$ in Eq.\eqref{eq:qm} and the $\zeta_m$ terms are determined with the $b_m$ being replaced by the $a_m$, see Eq.~\eqref{eq:zetam}. The reduced system of equations \eqref{3D_ODE} describes fully the leading-order droplet spreading dynamics as $\lambda\to 0$, confirming \textit{a posteriori} the assertion that both $|\dot{v}|$ and $U_m$   are  $O(1/|\ln\lambda|)$ as $\lambda\to0$. Like the 2D analysis, in the special case when
\begin{equation}\label{3D_ParabolicFlux}
	q(r,\phi,t) = \dot{v}(t)h(r,\phi,t)/v(t),
\end{equation}
the $\zeta_m$ terms vanish, so that terms involving $\dot{v}(t)$ are absent  from Eqs.~\eqref{3D_ODE} and volume changes appear only through the apparent contact angle, Eq.~\eqref{3D_ApparentContactAngle}.

The low-dimensional system \eqref{3D_ODE} may be simplified if further assumptions are invoked. For example, in Ref.~\cite{Savva2018}, where the case with $\dot{v}=0$ was treated, terms involving $\tilde{\beta}_m^0$, $\tilde{\beta}_m^\pm$ and $\psi_m$ with $m\ge1$ were omitted without compromising the generally favorable agreement reported with the solution to the governing PDE, since such terms do not arise from any logarithmically singular terms. Although in a consistent asymptotic scheme such terms need to be retained, we can argue heuristically that the  $O(\tilde{\beta}_m^0a_mU_0)$ terms make small contributions, because, by the time the contact line starts deforming, typically the axisymmetric spreading of the droplet nearly ceases so that $U_0\approx0$; the contributions of $O(\tilde{\beta}_m^\pm)$, just as those of  $O(U_1a_m)$ and $O(U_1^*a_m)$, may only become important for droplets with deformed contact lines which undergo long excursions away from their initial positions, whereas those of $\psi_m$ with $m\ge1$ merely introduce $O(\epsilon^2a_m)$ corrections to $U_m$. However, in all cases explored,  all the terms in Eqs.~\eqref{3D_ODE} were retained, since they may possibly accumulate non-negligible contributions when multiple periods of inflow/outflow are considered and because their inclusion adds insignificant computational overhead.

\section{Results}\label{3DSec_Results}

In this section we assess the outcomes of the asymptotic analysis of the preceding section through a series of numerical experiments which compare the predictions of Eqs.~\eqref{3D_ODE} with those of the governing equation and the appropriate initial conditions,  Eqs.~\eqref{BCs}, hereinafter referred to as the \emph{full model}. Central to the calculations using the asymptotic model, Eqs.~\eqref{3D_ODE}, is the way the apparent contact angle, $\vartheta$, is evaluated. Once $\vartheta$ is known, the system constitutes a set of generally weakly coupled and non-stiff integrodifferential equations and their implementation is relatively straightforward. It is decomposed into $2M+1$ unknowns for the evolution of $a_0$, $x_c$, $y_c$, and the real and imaginary parts of $a_m$, $m=2,3,\ldots,M$ (considering $M=50$ is typically sufficient, unless the structure of the heterogeneities has small-scale features). The whole process involves moving back and forth in Fourier space while time stepping is performed with a standard ordinary differential equation integrator, coupled with numerical quadrature for computing the time-dependent integrals  $\zeta_m(t)$ (see also Appendix.~\ref{sec:integral}).  

Just as in Ref.~\cite{Savva2018}, $\vartheta$ is computed using two different methods. The first is based on the perturbative  approximation of $\vartheta$ given by Eq.~\eqref{3D_ApparentContactAngle} and this model is referred to as \emph{reduced}.  Through the second approach, $\vartheta$ is determined more accurately for a given contact line shape, by solving Eqs.~\eqref{3D_LeadingOrderSystem} with the boundary integral method. In this manner, we formulate a linear integral equation which outputs $\vartheta$ by solving a linear system. Since this approach leverages the simpler system arising from the asymptotics, Eqs.~\eqref{3D_ODE}, with the numerical solution of leading-order problem, Eqs.~\eqref{3D_LeadingOrderSystem}, the resulting model is referred to as \emph{hybrid}. The general methodologies  adopted for solving the full, reduced and hybrid models, and for computing the associated integral terms are described in Ref.~\cite{Savva2018} (see Ref.~\cite{repo} for a Python implementation of the reduced and hybrid models).

Unless stated otherwise, we fix $a(\phi,0) = 1$, $\*x_c(0) = \*0$, and plot solutions to the full, hybrid, and reduced models with solid, dashed and dotted curves, respectively. Likewise, in all cases where $\theta(\*x)$ is represented by shading regions of the $x$-$y$ plane, darker/lighter shades correspond to larger/smaller contact angles. In some simulations we consider $p$-periodic volume variations governed by
\begin{equation}\label{3D_VolumeFunction}
	v(t) = \bar{v}+\frac{\tilde{v}}{\arctan 20}\arctan\left[\frac{20\sin(2\pi t/p)}{\sqrt{1+400\cos^2(2\pi t/p)}}\right],
\end{equation}
which corresponds to a nearly triangular wave with $\bar{v}-\tilde{v} \leq v(t) \leq \bar{v}+\tilde{v}$ (see, e.g., Fig.\ \ref{3DFig_SmallScale} and Part I). Throughout this work, we have fixed $\lambda=10^{-3}$, which is larger compared to the values considered in Part I, simply because the full problem becomes increasingly stiff as $\lambda\to0$, thus requiring significantly more computational resources to be simulated.  Also noteworthy is that, in most cases, the chemical heterogeneities $\theta(\*x)$ considered do not vary too sharply in order to avoid issues with retracting contact lines. In such cases, the contact line may develop protrusions which become elongated as the contact line retracts. Such features cannot be adequately resolved by the numerical scheme adopted and adaptive meshing techniques would be required. We chose not to pursue this route, since the validity of the analysis undertaken would be rather questionable in such cases.  

\subsection{Random substrates}

\begin{figure}[t!]
	\centering
	\includegraphics[scale = 1]{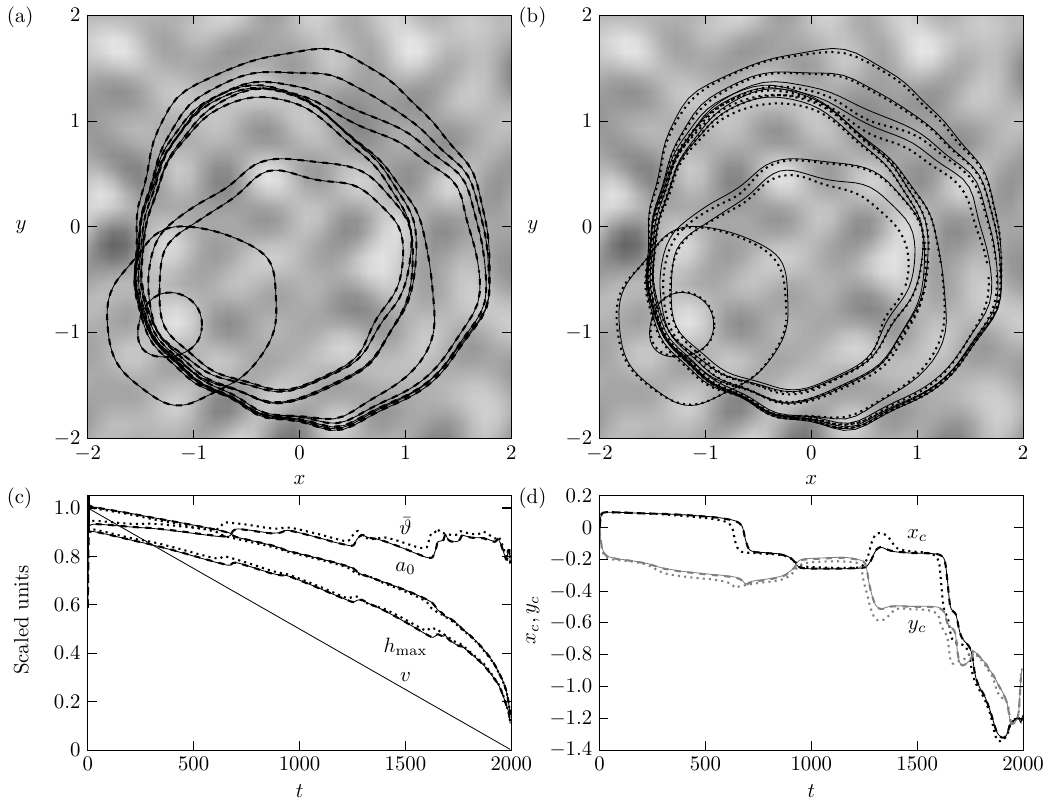}
	\caption{Stick--slip dynamics on a randomly varying $\theta(\*x)$ (see text for more details on how the profile is generated) with $v(t) = \pi\left(2-0.001t\right)$ and $q(\*x,t)$ prescribed according to Eq.~\eqref{3D_ParabolicFlux}. (a) and (b) show contact line profiles  when $t=10$, $200$, $400$, $600$, $800$, $1000$, $1200$, $1400$, $1600$, $1800$ and $1990$, comparing the solutions to the full problem with the hybrid and reduced problems, respectively (darker shades correspond to larger contact angles).  (c) Evolution of the mean radius $a_0$, mean macroscopic angle $\bar{\vartheta}$, maximum height $h_\text{max}$, and volume $v$, each scaled by $1.7$, $1.7$, $1.5$ and $2\pi$, respectively. (d) The evolution of the centroid coordinates, $x_c$ (black)  and $y_c$ (gray). In all plots, the solid, dashed and dotted curves correspond to solutions to the full, hybrid and reduced problems, respectively.}
	\label{3DFig_FourierLinearDecrease}
\end{figure}

The first example we consider is motivated by the experiments performed by \citeauthor{Dietrich2015} who investigated the evaporation of alcohol droplets in air and reported four evaporation modes, namely the constant contact angle and constant contact radius modes, as well as stick-slide and stick-jump modes, the latter of which is caused by the intermittent pinning of the contact line \cite{Dietrich2015}. The analysis presented in Sec.~\ref{3DSec_MatchedAsymptotics} does not account for evaporation, because the mass flux is maximized near the contact line \cite{Saxton2016,Savva2017,AJAEV2005}. However, one can qualitatively reproduce such a situation by considering a simple linear mass loss with Eq.~\eqref{3D_ParabolicFlux} for the distribution of the mass transfer. To induce a non-trivial centroid motion, we prescribe noisy chemical heterogeneities of the form
\begin{equation}\label{3D_RandomSubs}
	\theta(\*x) = \hat{\theta}(\*x)+\tilde{\theta}(\*x),
\end{equation} 
where  $\hat{\theta}(\*x)$ is a deterministic part giving the dominant structure of the substrate, and $\tilde{\theta}(\*x)$ corresponds to a realization of spatial band-limited white noise. In Part I of this study, similar substrates were considered for exploring the snapping dynamics observed by \cauthor{Wells2018} (see Fig.\ 10 in Part I).

Figure~\ref{3DFig_FourierLinearDecrease} shows a case of a substrate with heterogeneities of the form \eqref{3D_RandomSubs}, where $\hat{\theta}(\*x) \equiv 1.5$ and $\tilde{\theta}(\*x)$ is represented by a superposition of $75$ harmonics with wavenumbers up to $3\pi$ and whose amplitudes are normally distributed with zero mean and unit variance, ensuring that $\theta(\*x)>0$ everywhere. By tracking the mean values of $a(\phi,t)$, $a_0$, and the apparent contact angle $\vartheta$, $\bar{\vartheta}$, we observe behaviors that are qualitatively consistent with the observations recorded in Fig.~2 of Ref.~\cite{Dietrich2015}, noting that we used $h_\text{max} = a_0\bar{\vartheta}/2$ as a proxy for the maximum height of the droplet, found by setting $r=0$ in the leading term of Eq.~\eqref{3D_LeadingOrderOuter}. Although the corresponding measurements in experiments are usually taken from photographs of the side of the droplet, recording the mean values of $a$ and $\vartheta$ from nearly circular contact lines as done here gives similar results (see the contact line profiles in Fig.\ \ref{3DFig_FourierLinearDecrease}(a)).

As the droplet loses mass, its mean radius exhibits a series of jumps which is also marked with a temporary increase in $h_\text{max}$ and $\bar{\vartheta}$. Usually in these circumstances the contact line will locally remain pinned at a lower wettability region, thus causing other parts of the contact line to respond to this pinning by shifting towards the pinning site as the droplet loses mass (see Figs.~\ref{3DFig_FourierLinearDecrease}(a), (b) and (d)). Once the contact line manages to overcome a wettability barrier, it exhibits a stick-jump event. These behaviors corroborate the experimental observations of \cauthor{Dietrich2015} which were also attributed to spatial variations in surface heterogeneities. Noteworthy also is the excellent agreement between the predictions offered by the full and hybrid models which are nearly indistinguishable. The reduced model on the other hand shows some disagreement, but captures the generic features reasonably well.

\begin{figure}[t!]
	\centering
	\includegraphics[scale = 1]{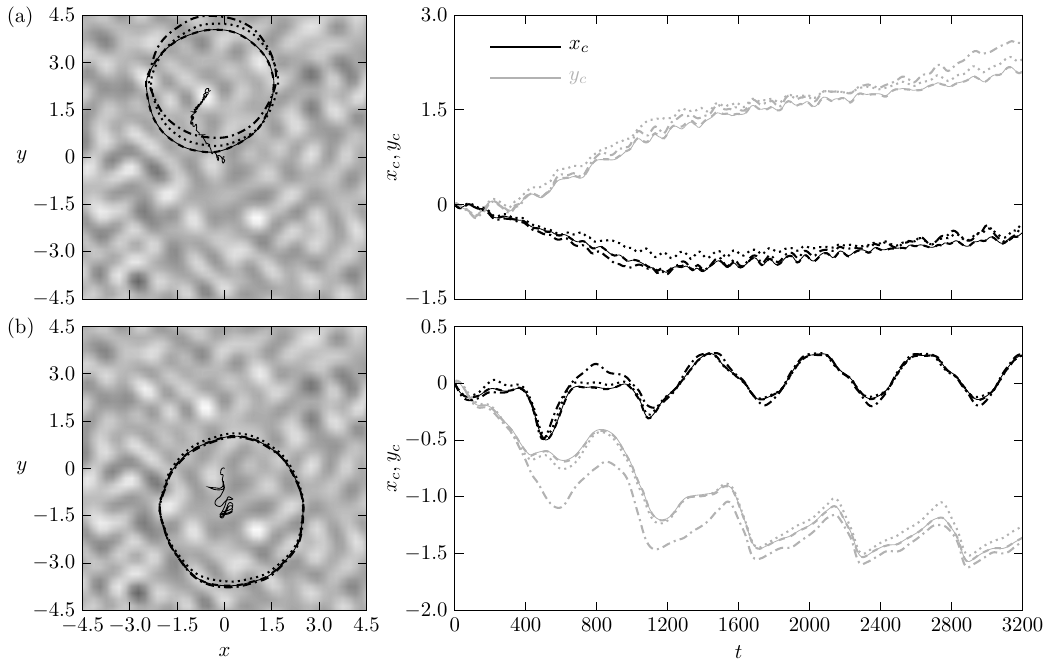}
	\caption{Periodic variations in droplet volume over a substrate comprised of randomized heterogeneous features (see text for a brief description on how the substrate is generated), where $q(\*x,t)$ is prescribed according to Eq.~\eqref{3D_ParabolicFlux} and $v(t)$ according to Eq.~\eqref{3D_VolumeFunction} with $\bar{v} = 2\pi$, $\tilde{v} = 1.5\pi$. In panel (a), $p=200$; in panel (b), $p=600$. In each panel, the left plot shows the contact line shape at $t=3200$ for the full model (solid curves), the hybrid model using Eqs.~\eqref{3D_ODE} (dashed curves), the reduced model using Eqs.~\eqref{3D_ODE} (dotted curves), and the hybrid model using the leading-order Eq.~\eqref{Lacey}  (dash dotted curves), and the path followed by $\*x_c(t)$; the right plot shows the corresponding evolution of the centroid coordinates $x_c(t)$ (black curves) and $y_c(t)$ (gray curves).}
	\label{3DFig_MultipleFlowFourier}
\end{figure}

One of the key observations of Part I of the present study was that periodic variations in the droplet mass led to periodic 2D droplet dynamics after the decay of initial transients, which typically occur within three to four periods. To assess how well this applies in the 3D setting, we use a substrate with random heterogeneities prescribed according to Eq.~\eqref{3D_RandomSubs} with $\hat{\theta}(\*x) = 1$ and where $\tilde{\theta}(\*x)$ is a superposition of $10$ harmonics with wavenumbers up to $2\pi$, whose amplitudes are normally distributed with zero mean and variance set to $0.3$. For such a substrate, the chemical heterogeneities vary more weakly compared to the substrate used in Fig.~\ref{3DFig_FourierLinearDecrease}, giving rise to softer pinning transitions which almost entirely eliminate stick--slip dynamics. For this choice for $\theta(\*x)$ the contact line is more circular, thus requiring fewer collocation points in the azimuthal direction to be accurately resolved and permitting efficient simulations for far larger values of $t$.

The outcome is shown in Fig.~\ref{3DFig_MultipleFlowFourier},  plotting how the centroid evolves for two different values of the period of mass fluctuations, namely $p=200$ and $p=600$. Remarkably, the droplets are driven to entirely different locations on the substrate. Moreover, the dynamics when $p=200$ does not appear to have settled to a periodic state within the simulated time.  Hence, this result points to the possibility of quasiperiodic dynamics, induced by the non-linear coupling of the random features and the period of inflow/outflow. However, it is expected that larger values of $p$ (e.g., see the case when $p=600$ in Fig.~\ref{3DFig_MultipleFlowFourier}) and/or weaker heterogeneities will mitigate this effect, facilitating the transition to periodic motion as in the 2D case. Lastly, we also need to highlight the importance of including the correction terms derived from the asymptotic analysis undertaken, by performing the same set of simulations with the hybrid approach, but using only the leading-order term of the asymptotic theory \cite{Lacey1982}, namely
\begin{equation}\label{Lacey}
	\partial_t\*c\cdot\*\nu = \frac{\theta_*^3-\vartheta^3}{3\ln\lambda},
\end{equation}
which only captures the  $O(1/|\ln\lambda|)$ terms of Eq.~\eqref{3D_ODE}, which dominate as $\lambda\to0$. As seen in the plots of Fig.~\ref{3DFig_MultipleFlowFourier}, just the leading-order term is insufficient, at least for the sizes of $\lambda$ considered, since we observe appreciable departures from the simulations with the full model. At the same time, the computations with the reduced model, which uses Eq.~\eqref{3D_ApparentContactAngle} for $\vartheta$, also highlight that it is equally important to accurately compute $\vartheta$ for improved agreement. Without these additional terms, which are better captured with the hybrid method applied to Eqs.~\eqref{3D_ODE}, we see that heterogeneities may occasionally mistime stick--slip events, thus moving the droplet to a different location.


\subsection{Localized mass transfer}

\begin{figure}
	\centering
	\includegraphics[scale = 1]{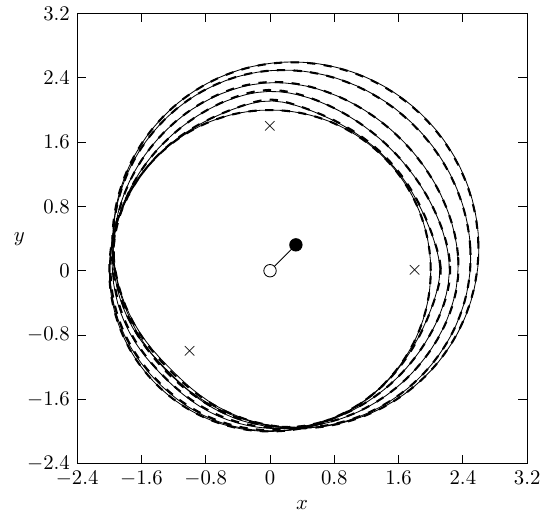}
	\caption{Localized mass transfer on a homogeneous substrate $\theta(\*x) = 1$, by solving Eqs.~\eqref{GovPDE} (solid curves) and Eqs.~\eqref{3D_ODE} with the hybrid method (dashed curves), when $\dot{q}=\dot{v}\left[ \delta(x-1.8)\delta(y) + \delta(x)\delta(y-1.8)-\delta(x+1)\delta(y+1) \right]$ (location of fluxes are shown with crosses).  Volume changes are governed by Eq.~\eqref{3D_VolumeTanh} with $v_0 = 2\pi$, $v_\infty=3\pi$ and $\eta=1/30$. The initial contact line satisfies $a(\phi,0)=2$, so that the droplet is at equilibrium for the starting volume $v(0)=2\pi$. Droplet profiles shown correspond to times $t=0$, $2$, $8$,  $16$, $32$ and $60$. The centroid is displaced from the origin (open circle) along the line $y=x$ to a new position, marked with a solid circle when $t=60$.}
	\label{3DFig_LocalCLShapes}
\end{figure}

\begin{figure}
	\centering
	\includegraphics[scale = 1]{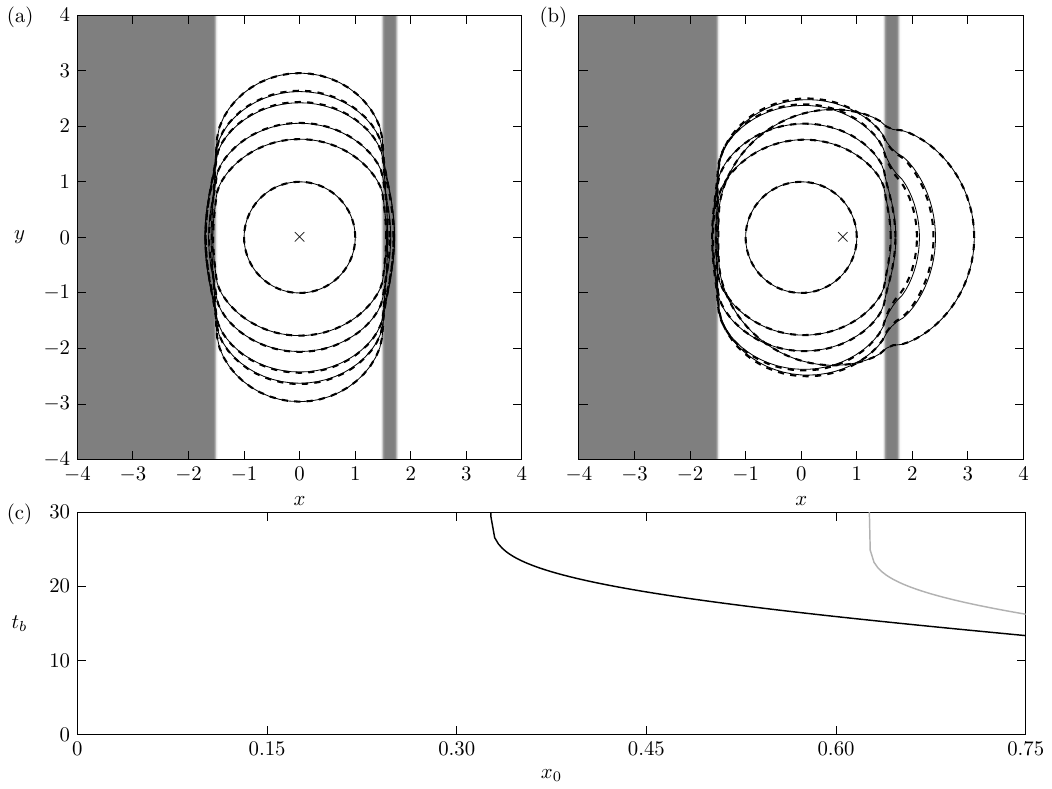}
	\caption{Changing the localization of the mass transfer from $(x_0,y_0) = (0,0)$ to $(0.75,0)$ with the heterogeneous barriers $\theta(\*x) = 1.2+\tilde{g}$ $\{\tanh\left[50(x-1.5)\right]$ $-\tanh\left[50(x+1.5)\right]-\tanh\left[50(x-1.75)\right]\}$. Plots (a) and (b) show droplet profiles at times $t = 0$, $5$, $10$, $20$, $30$ and $300$ for localization at $x_0 = 0$ and $x_0 = 0.75$, respectively (localization depicted by crosses). The substrate in (a) and (b) is shaded according to the choice of $\theta$ where dark/light patches correspond to $\theta \approx 1.2\pm 0.25$. (c) The time at which the point $x_c(t)+a(0,t)$ breaks the heterogeneous barrier, $t_b$, against $x_0$, where black and gray curves are for $\tilde{g} = 0.25$ and $\tilde{g} = 0.275$, respectively. In all plots volume changes are given by \eqref{3D_VolumeTanh} with $v_0 = \pi$, $v_\infty=3\pi$ and $\eta=\pi/50$ (PDE computations in (a) and (b) are carried using Eq.~\eqref{3D_LocalisedFlux} with $S=20$).}
	\label{3DFig_LocalWallBreak}
\end{figure}

The analysis presented is capable of capturing localized mass transfer provided that $q$ is non-zero only within the droplet footprint, $\Omega(t)$. Localised forms of $q$ have also been explored in Part I, showing that they may reposition the droplet at a different location on the substrate, or even induce topological transitions through break-up. It is thus of interest to explore these effects in the 3D setting as well, by considering a Gaussian flux distribution of the form
\begin{equation}\label{3D_LocalisedFlux}
	q = \frac{\ds \dot{v}(t)\exp\left\{-S\left[(x-x_0)^2+(y-y_0)^2\right]\right\}}{\ds \int_{\Omega(t)}\exp\left\{-S\left[(x-x_0)^2+(y-y_0)^2\right]\right\}\,\mathrm{d}\*x},
\end{equation}
whose peak is located at $(x_0,y_0)\in \Omega(t)$ with $S\gg1$ so that we have $q\approx0$ when $\*x=\*c$, noting that the denominator is estimated with numerical quadrature. In order to make computations with Eq.~\eqref{GovPDE} feasible, we only consider moderate values of $S$, chosen as a compromise between the requirement for the analysis to hold that $q$ vanishes along the contact line, while capturing its dynamics with fewer collocation points than those that would have been required to accurately resolve $q(\*x,t)$ for much larger values of $S$. On the other hand, the implementation efficiency of Eqs.~\eqref{3D_ODE} is enhanced if we formally take the limit $S\to\infty$, so that Eq. \eqref{3D_LocalisedFlux} reduces to $q=\dot{v}\delta(x-x_0)\delta(y-y_0)$, where $\delta(x)$ is the Dirac delta function. This is because, in this limit, we avoid having to compute the Fourier decomposition of $q$ at each time step. In the moving frame,   $q=\dot{v}\delta(x-x_0)\delta(y-y_0)$ may be cast in the form
\refstepcounter{equation}
\begin{equation*}\label{eq:deltar}
	q = \frac{\dot{v}(t)\delta(r-r_0)\delta(\phi-\phi_0)}{ra(\phi,t)^2},\eqno{(\theequation{\text{a}})}
\end{equation*}
where
\begin{equation*}
	\phi_0 = \tan^{-1}\left(\frac{y_0-y_c}{x_0-x_c}\right)\quad\text{and}\quad r_0 = \frac{\sqrt{(x_0-x_c)^2+(y_0-y_c)^2}}{a(\phi_0,t)},\eqno{(\theequation{\text{b,c}})}
\end{equation*}
with $r_0<1$. Equations~\eqref{eq:deltar} give the location of the mass transfer in the $(r,\phi)$ computational coordinate system so that the injection point remains unchanged in the physical coordinate system. This allows us to evaluate the time-dependent integrals $\zeta_m(t)$ to obtain (see Eq.~\eqref{eq:zetam})
\begin{subequations}\label{Delta_Integrals}
	\begin{align}
		\zeta_0(t) &= \frac{r_0^2}{\pi a^2(\phi_0,t)\vartheta(\phi_0,t)(1-r_0^2)}-\frac{a_0}{4v(t)},\\
		\zeta_m(t) &= \frac{f_m(r_0)\e^{-\ii m\phi_0}}{\pi a^2(\phi_0,t)\vartheta(\phi_0,t)}-\frac{a_m}{v(t)}\int_0^1 f_m(r)r\left[ r^m-1+\frac{m+1}{2}\left( 1-r^2\right)\right]\,\mathrm{d}r, \quad m\geq 1,
	\end{align}
\end{subequations}
which may be further simplified by using the approximation $a^2(\phi_0,t)\vartheta(\phi_0,t)\approx 4v(t)/[\pi a_0(t)]$. The integral term in Eq.~(\ref{Delta_Integrals}b) can be pre-computed with numerical quadrature, and the value of $f_m(r_0)$ is easily determined with spectrally accurate polynomial interpolation. In the examples that follow, and whenever comparisons with the governing PDE are made, we use Eq.~\eqref{3D_LocalisedFlux} in Eq.~\eqref{GovPDE}, but in simulations of  Eqs.~\eqref{3D_ODE} we use Eqs.~\eqref{eq:deltar}  and \eqref{Delta_Integrals}, not expecting appreciable differences especially when the flux is not too close to the contact line. We also assume that the volume varies according to
\begin{equation}\label{3D_VolumeTanh}
	v(t) = v_0+(v_\infty-v_0)\tanh\left(\eta t\right),
\end{equation}
which monotonically increases from $v=v_0$ at $t=0$ to $v\to v_\infty$ as $t\to\infty$, where $\eta$ is a generally small parameter to ensure a slow transition to the final volume in accordance with the assumptions of the theory.

\begin{figure}
	\centering
	\includegraphics[scale = 1]{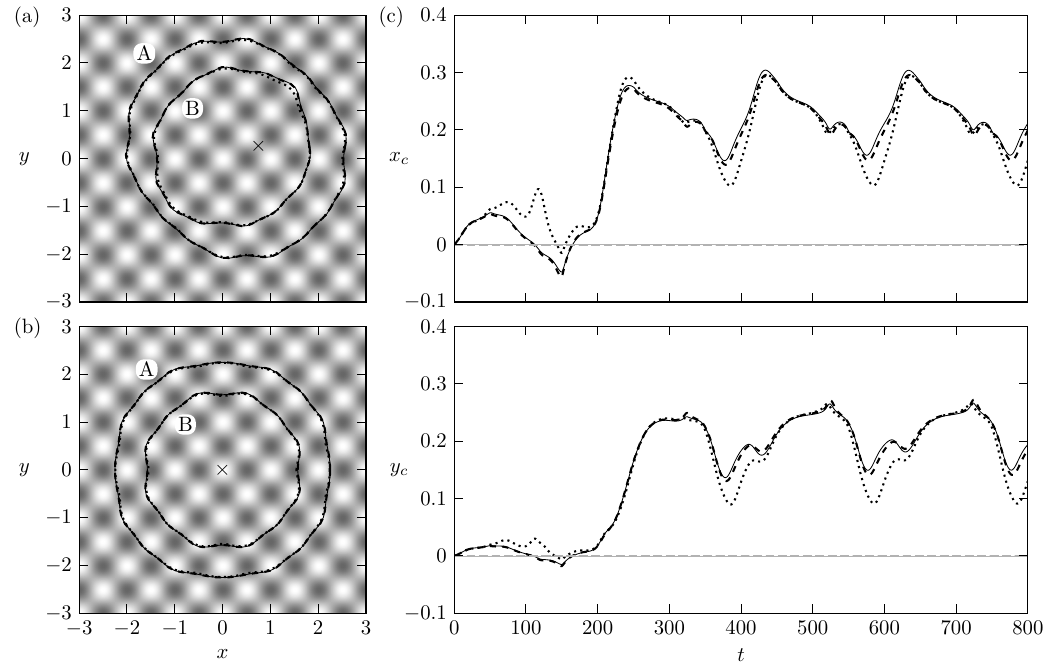}
	\caption{Transition to periodic dynamics for  the substrate with $\theta(\*x) = 1-0.15\left\{\cos\left[2\pi(x+y)\right]+\cos\left[2\pi(x-y)\right]\right\}$. Contact line shapes for the full, hybrid and reduced models at times $t=650$ (marked with `A') and $t=750$ (marked with `B') when (a) $(x_0,y_0)=(0.75,0.25)$ and (b) $(x_0,y_0)=(0,0)$; the source/sink locations are marked with an $\times$. The contact line shapes `A' and `B' correspond to the shapes at the maximum/minimum volumes after the droplet settles to a periodic state. Panel (c) shows the evolution of the coordinates of the centroid $x_c(t)$ (top plot) and $y_c(t)$ (bottom plot) when $(x_0,y_0)=(0.75,0.25)$ (black curves) and $(x_0,y_0)=(0,0)$ (gray curves). For both cases, the volume varies according to Eq.~\eqref{3D_VolumeFunction} with $\bar{v} = 2\pi$, $\tilde{v} = \pi$ and $p = 200$; for the PDE solution Eq.~\eqref{3D_LocalisedFlux} is used with $S=20$; for the reduced and hybrid models Eqs.~\eqref{eq:deltar} have been used.}
	\label{3DFig_LocalisedMultFlow}
\end{figure}

In Fig.~\ref{3DFig_LocalCLShapes} we consider a droplet which is initially centered at the origin, has volume $v(0)=v_0=2\pi$ and is at equilibrium on a homogeneous substrate with $\theta(\*x)\equiv1$, i.e. $a(\phi,0)=2$. Liquid is added into the droplet according to Eq.~\eqref{3D_VolumeTanh} with $v_\infty=3\pi$  and $\eta=1/30$ using 2 injection points located close to the contact line at $(1.8,0)$ and $(0,1.8)$ and one withdrawal point at $(-1,-1)$, so that $\dot{q}=\dot{v}\left[ \delta(x-1.8)\delta(y) + \delta(x)\delta(y-1.8)-\delta(x+1)\delta(y+1) \right]$. For solving the PDE, the delta functions are replaced with the appropriate scaled Gaussians using Eq.~\eqref{3D_LocalisedFlux} with $S=100$. The localized injection points contribute to the development of two protrusions in the contact line along the positive $x$- and $y$-axes, which are more pronounced during the early stages. Ultimately, however, the contact line relaxes to a circular shape in the long-time limit, as expected, with the centroid being displaced slightly along the line $y=x$.

As already shown in Part I, the interplay of substrate heterogeneities and the way changes in mass occur is rather intricate and small changes to either of them may cause a markedly different behavior. This kind of subtle interplay can be leveraged in applications as a means to control droplet transport, e.g.\ for sorting droplets of different sizes~(see, e.g.\ Ref.~\cite{Kusumaatmaja2007}; also Ref~\cite{Xi2017} for a review on droplet sorting). Another situation of interest is using mass transfer to assist a droplet in escaping nearby wettability barriers. This is highlighted in Fig.~\ref{3DFig_LocalWallBreak}, where we consider a droplet confined between two parallel heterogeneous stripes of lower wettability. Using $q$ as prescribed by Eq.~\eqref{3D_LocalisedFlux} and $v$ by Eq.~\eqref{3D_VolumeFunction}, we alter the position of the injection point $(x_0,0)$ for $x_0\ge0$ to determine locations which allow the droplet to escape the rightmost wettability barrier. When $x_0$ is small, the droplet remains trapped between the two heterogeneous barriers, see e.g.\ Fig.\ \ref{3DFig_LocalWallBreak}(a). This type of behavior has been reported in related works with droplets of constant mass~\cite{Bliznyuk2009,David2012,Damle2017}. When $x_0$ becomes sufficiently large, the droplet may eventually overcome the rightmost heterogeneous stripe (see Fig.~\ref{3DFig_LocalWallBreak}(b)). These two distinct states are highly dependent both on the value of $x_0$ and the wettability contrast, see Fig.~\ref{3DFig_LocalWallBreak}(c). In fact, altering the (non-dimensional) difference in the local contact angles between the stripes from $0.5$ ($\tilde{g}=0.25$) to $0.53$ ($\tilde{g}=0.265$) causes almost a twofold increase in the minimum value of $x_0$ required to break the barrier, from about $0.33$ to about $0.63$. It is also worth noting that the calculations in  Fig.~\ref{3DFig_LocalWallBreak}(c) were performed with the hybrid model only, where over 100 hybrid-model simulations required far less time to complete than a single full-model simulation.

A feature which persists with the localized mass transfer cases explored is that changing the mass near the contact line can move the droplet preferentially in one direction, sometimes against heterogeneous barriers if these are sufficiently weak, or if the mass transfer is sufficiently strong (see Fig.\ \ref{3DFig_LocalWallBreak}). Therefore, mass variations may be used as a mechanism to overcome the energy barriers introduced by chemical heterogeneities. To demonstrate this plausibility,  a heterogeneity profile comprising of periodically varying heterogeneous features is considered in Fig.~\ref{3DFig_LocalisedMultFlow}. By coupling a localized form of $q$ with periodic mass changes, we observe, like in Part I, that the droplet attempts to center itself around the inlet/outlet position of the fluid transfer (see also Fig.\ 7, Part I). However, the presence of heterogeneities may prevent this from happening if these are sufficiently strong (see Fig.~\ref{3DFig_LocalisedMultFlow}(a)). Let us also note that, as expected, and in agreement with the observations of Part I, the dynamics ultimately become periodic in the long-time limit as a result of coupling periodic flow conditions and heterogeneous features. At times, however, we temporarily observe slight departures between the hybrid and full models which are usually exacerbated when  $q$ is about to switch from inflow to outflow and vice  versa (this issue is more pronounced for the reduced model). In such cases, the asymptotic model is inapplicable and errors typically accumulate more quickly for smaller values of $p$. As mentioned previously, as well as in Part I, the asymptotic analysis applies when $\lambda\ll|\pd[a]{t}|\ll1$ and, strictly speaking, it is rendered invalid when the fronts are about to switch their direction of motion. This, however, does not always occur and it strongly depends on the combined effect of the forcing through $q$ and the surface heterogeneities, $\theta(\*x)$.

\subsection{Small-scale features}
\abovecaptionskip=0pt
\begin{figure}[t!]
	\centering
	\includegraphics[scale = 0.95]{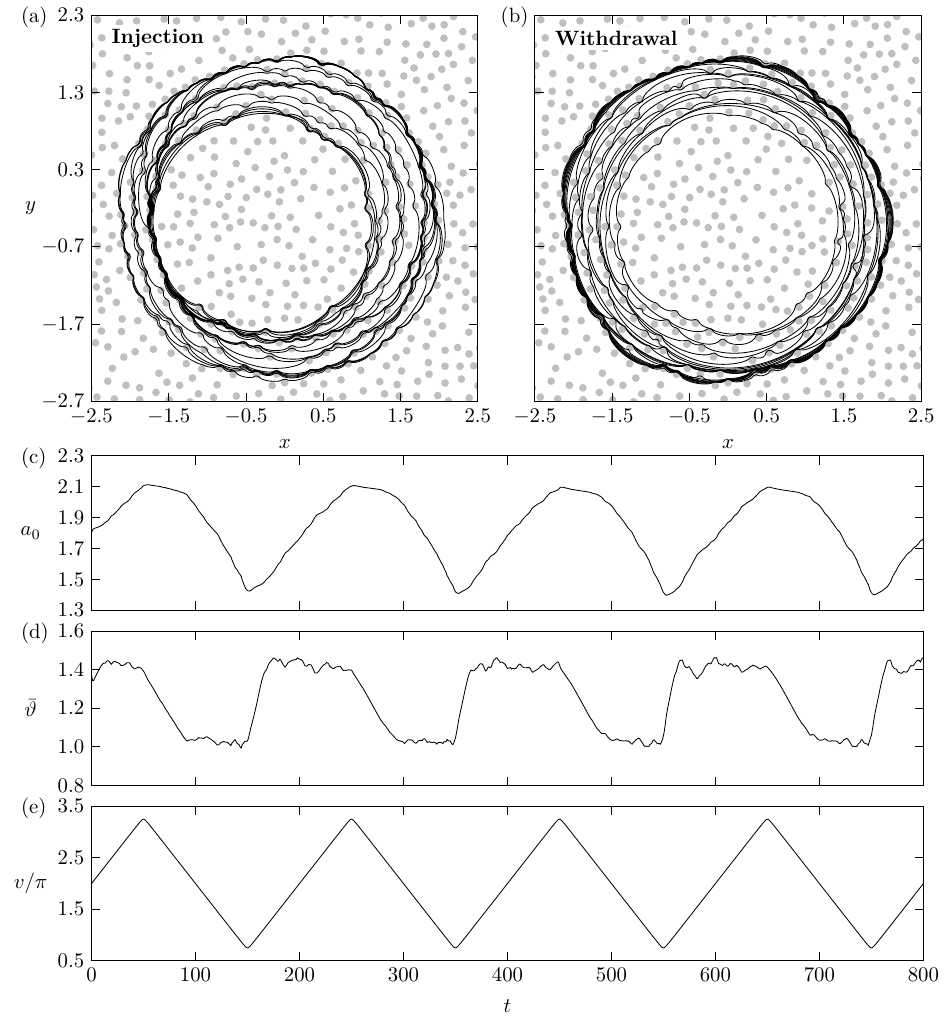}
	\caption{Volume cycling over a substrate with randomly placed small patches of lower wettability using the reduced model (see text for details on how the substrate is generated). The volume varies according to Eq.~\eqref{3D_VolumeFunction} with parameters $\bar{v}=2\pi$, $\tilde{v}=1.25\pi$ and $p=200$, together with $q(\*x,t)=\dot{v}(t)\delta(\*x)$.  Droplet profiles during (a) fluid injection when $350\le t\le450$; (b) fluid withdrawal when $450\le t\le550$. For both cases, profiles are drawn every 5 time units. (c), (d) and (e) show, respectively, the evolution of the the mean radius, $a_0$, mean apparent contact angle, $\bar{\vartheta}$, and the volume, $v$, scaled by $\pi$.}
	\label{3DFig_SmallScale}
\end{figure}

In this final example we consider a substrate formed with randomly distributed heterogeneous features by using
\begin{equation}\label{3D_RandomSmallScale}
	\vartheta(\*x) = 1+\sum_{j=1}^{800}\tilde{\theta}\left(\sqrt{\left(x-\bar{x}_j\right)^2+\left(y-\bar{y}_j\right)^2}\right),\quad \tilde{\theta}(x) = \frac{1}{2}\left[\tanh\left(200x+10\right)-\tanh\left(200x-10\right)\right], 
\end{equation}
which prescribes $800$ localized circular patches of lower wettability ($\vartheta\approx 2$) with approximate radius $0.05$ units and positions $(\bar{x}_j,\bar{y}_j)$ chosen randomly to lie in the square $(-3,3)\times (-3,3)$, so that the features are at least $0.15$ dimensionless units apart. A similar calculation has been performed in Ref.~\cite{Savva2018} for the case of constant mass to examine stick-slip behaviours with advancing contact lines. Here, we are interested in the dynamic phenomena that arise from cycling the volume through periodic mass changes. To properly resolve contact line variations occurring at shorter scales would require a very large number of collocation points in the azimuthal direction, which makes solving the full model inefficient with our current implementation and the hybrid model rather challenging due to the solution of large dense matrix-vector equations at each time-step. Hence, we chose to perform this simulation with the reduced model only, although strictly speaking we no longer have $a_0\gg\aphi$ as required by the asymptotic analysis undertaken. However, since the reduced model works generally well for other cases considered, we can use it to gather a qualitative outlook on the dynamics that arise  and to demonstrate that the model is able to capture the generic features that arise in experimental settings rather well.

The result of the computation is shown in Fig.\ \ref{3DFig_SmallScale}, where dynamically cycling the droplet volume with $q(\*x,t)=\dot{v}(t)\delta(\*x)$ gives rise to a number of features present in previous examples. Firstly, it is easy to see from Figs.\ \ref{3DFig_SmallScale}(a) and (b) that pinning/de-pinning events emerge throughout both the stages where fluid is injected and withdrawn. The constant-radius and angle modes also appear as a consequence of the pinning events which suggests, like in Part I, that such behaviours may arise in experimental settings due to substrate features whose effects are more difficult to quantify (see Figs.~\ref{3DFig_SmallScale}(c) and (d)). Interestingly, the dynamics presented  here are qualitatively very similar to Fig.\ 10 in Ref.~\cite{Lam2002} by \citeauthor{Lam2002}, where the authors experimentally analyze cycling the droplet's volume through a needle at its base. Like the results here, \citeauthor{Lam2002} showed that the constant-radius mode is rather brief (evidenced, e.g., by the clustering of contact line profiles in Fig.~\ref{3DFig_SmallScale}b when $t\approx450$) and occurs shortly after the flow conditions switch, giving predominately the constant-angle mode. Noteworthy also is that the stick-slip and pinning behaviours are reminiscent of the results in Refs.\ \cite{Cubaud2004} and \cite{Cubaud2001} by  Cubaud \emph{et al.}, although no direct comparisons can be made because larger droplets are considered where gravitational effects become appreciable, which, in the present case, are neglected in favor of analytical tractability.

\section{Concluding Remarks}\label{3DSec_Conclusion}

The motion of a liquid drop on a solid surface is a process that is rather easy to conceptualise, with numerous applications across science and engineering. However, the underlying physical processes are inherently complex, rendering their study highly non-trivial. Although the long-wave limit yields a single evolution equation for the droplet thickness, its numerical solution remains considerably challenging particularly as $\lambda\to0$. The asymptotic analysis undertaken in this two-part study, is able to mitigate this challenge by developing non-stiff lower-dimensional models for the evolution of the contact line in the 2D and 3D settings.

In this part, we tackled the 3D geometry, building upon the work in Ref.~\cite{Savva2018} to include the additional terms required to capture droplets of variable mass as prescribed through some spatiotemporal flux function. To simplify the analysis based on the observations of Part I, the assumption that the mass transfer vanishes along the contact line was applied so that explicit equations could be obtained for the Fourier coefficients of the contact line, see Eqs.~\eqref{3D_ODE}. In these equations, like their analogous 2D counterpart (see Eqs.\ (17), Part I), the contributions of the rate of change of the droplet volume on the contact line shape were accounted for by incorporating time-dependent integral terms involving the mass transfer. To assess the validity of the theory presented here and to explore some interesting physical phenomena associated with droplet spreading, we discussed a number of representative cases that contrast our theory with simulations to the governing system, Eq.~\eqref{GovPDE}. In the all of the cases considered, we observed excellent agreement between the outcomes of the analysis and the numerical solutions of the full equations, noting that the most optimal agreement is obtained with the hybrid model, particularly in cases when the contact line is more strongly deformed by heterogeneities.

The competition between mass transfer and chemical heterogeneity was examined by considering a number of cases motivated by experiments using two specific forms for the mass flux, namely Eqs.~\eqref{3D_ParabolicFlux} and \eqref{3D_LocalisedFlux}, and its limiting form, Eqs.~\eqref{eq:deltar}. However, other forms of $q$ could have been used as well, which, when combined with the chemical heterogeneities result in a rich dynamical system. Appropriately tuning $q$ and $\theta$ allows us to control droplet transport and confinement. For example, this may be achieved by introducing wettability barriers and chemical gradients in conjunction with other forms for $q$ or with localized fluxes which are positioned sufficiently close to the contact line as a means to push or pull the droplet in the desired direction. The influence of these competing mechanisms deserve a more detailed investigation, which is beyond the scope of the present work. Let us also note that no quantitative comparison with experiments was sought, since, as far as we are aware, no study in the literature reports the time evolution of the contact line, at least within the regime of applicability of our theory. However, we were able to demonstrate that features commonly observed in experimental settings naturally arose through simulations, including stick-slide/stick-jump events and the constant-radius and constant-angle modes. Notably, qualitative comparison between the experimental studies of \citeauthor{Dietrich2015} who consider evaporating droplets (see Fig.~\ref{3DFig_FourierLinearDecrease}) \cite{Dietrich2015}, and \citeauthor{Lam2002} who consider liquid injection/withdrawal with a needle (see Fig.\ \ref{3DFig_SmallScale}) \cite{Lam2002}, were demonstrated. These comparisons were performed with substrates decorated with random heterogeneities. Thus, the observed behaviors emerge due to substrate heterogeneities, whose effects are generally more difficult to assess both in experiments and in direct numerical simulations. Just like the 2D results, we noted that the dynamics is quite sensitive to the parameters controlling the mass transfer and chemical heterogeneity, showing how the subsequent behaviors can be drastically changed by slightly altering them. Nevertheless, Eqs.~\eqref{3D_ODE} were able to capture these behaviors and were in excellent agreement with the solution to the governing PDE. Hence, the effort invested in developing the two-term asymptotic expansion of the solution is rewarded with a more favorable agreement compared to solutions obtained using the leading-order equation, with little additional computational overhead, see, e.g., Fig.~\ref{3DFig_MultipleFlowFourier} and compare Eqs.~\eqref{eq:CLlaw} and~\eqref{Lacey}. Although the discussion focused on the inverse linear slip model, it should be emphasized that exactly the same equations apply for the Navier slip model, and other contact line models may be invoked through a relatively straightforward rescaling of $\lambda$ to map to its corresponding microscopic lengthscale \cite{Savva2011,Sibley2014}.

Throughout this two-part study, the assumption that the mass transfer vanishes at the contact line was applied so that changes in mass occurred through the bulk of the droplet.  This assumption was not deemed to be too restrictive for our purposes, and, indeed, it did not appear to impact the features of the phenomenology we wished to uncover. However, it is inappropriate for investigating mass loss through evaporation, since, in that case, the mass flux is maximized close to the contact line. The asymptotics of such cases, require  a specialized treatment of the microscale \cite{Saxton2016,Savva2017}, a subject to be explored in more detail in a forthcoming investigation.

\section{Acknowledgements}
The computations of this work were performed using the computational facilities of the Advanced Research Computing @ Cardiff Division, Cardiff University. DG acknowledges support from the Engineering and Physical Sciences Research Council of the UK through Grant No EP/P505453/1; NS acknowledges support from the European Union's Horizon 2020 research and innovation programme under grant agreement No 810660.

\appendix

\section{Local expansion of the governing PDE}\label{sec:BC}

The equation for the evolution of the two-dimensional contact line $\*c(\*x,t)$, Eq. (\ref{GovPDE}d), is formally derived by following similar considerations which were presented in related works \cite{Young1994,OLIVER2015,Savva2018}, namely by taking a local expansion of the governing PDE
\begin{equation}\label{Append_GovPDE}
	\pd[h]{t}+\Nabla\cdot\left[h\*Q\right] = q
\end{equation}
near $\*x=\*c$. Here we let $\*Q = \left(h^2+\lambda^2\right)\Nabla\nabla^2 h$,  noting that $\*Q$ would change should a different slip model be used, or if additional complexities, such as substrate topography, be included. The leading term of the series expansion near the contact line is
\begin{equation}
	h = (\*x-\*c)\cdot\Nabla h|_{\*x=\*c}+\ldots,
\end{equation}
where dots denote omitted higher-order corrections, which vanish as $\*x\to\*c$. Using this expansion, we can deduce that
\begin{equation}
	\partial_t h|_{\*x=\*c} = -\partial_t \*c\cdot\Nabla h|_{\*x=\*c}+\ldots,
\end{equation}
which, when combined with an expanded form of Eq.~\eqref{Append_GovPDE} about $\*x=\*c$, yields
\begin{equation}
	-\pd[\*c]{t}\cdot\Nabla h|_{\*x=\*c}+\*Q|_{\*x=\*c}\cdot\Nabla h|_{\*x=\*c}+\ldots = q|_{\*x=\*c}.
\end{equation}
This implies that
\begin{equation}\label{Mov_Bound_1}
	\left[(\*Q-\partial_t\*c)\cdot\Nabla h\right]|_{\*x=\*c} = q|_{\*x=\*c},
\end{equation}
which is the corresponding moving boundary condition. This expression can be further simplified using
\begin{equation}
	\Nabla h|_{\*x=\*c} = (\Nabla h\cdot\*\nu)|_{\*x=\*c}\*\nu+(\Nabla h \cdot \*\tau)|_{\*x=\*c}\*\tau,
\end{equation}
where $\*\tau$ is the unit tangent vector to the contact line (see Fig.\ \ref{3DFig_CoordTransform}). Since $h=0$ on $\*c$, $ \left.(\*\nabla h \cdot\*\tau)\right|_{\*x=\*c}=0$ so that
\begin{equation}\label{Mov_Bound_2}
	\Nabla h|_{\*x=\*c} = (\Nabla h \cdot \*\nu)|_{\*x=\*c}\*\nu = -\theta|_{\*x=\*c}\*\nu.
\end{equation}
Therefore, using Eq.~\eqref{Mov_Bound_2} in Eq.~\eqref{Mov_Bound_1}, and evaluating $\*Q$ at $\*x=\*c$  gives the moving boundary condition, Eq. (\ref{GovPDE}d).

\section{The correction term of the leading-order outer solution}\label{sec:BVPs}
Here we outline the derivation of the correction to the leading-order outer solution. Starting from Eq.~\eqref{eq:h1q}, write
\begin{equation}
\nabla^2h_1 = \frac{1}{a\vartheta^2}\left(p_0 + \delta p_1+\ldots\right),\label{eq:Laph1}
\end{equation}
where
\begin{align}
p_0 &= \mathcal{L}_0 (h_{1,0})+\mathcal{L}_1 (h_{1,1})\e^{\ii\phi}, \\
p_1 &= \sum_{m=1}^{M}\left\{\mathcal{L}_m(\tilde{h}_{1,m}) + m^2\frac{b_m}{b_0} \mathcal{K}_m(h_{1,0})+\frac{m^2-1}{2}\left[\frac{b_{m+1}}{b_0} \mathcal{K}_{m+1}(h^*_{1,1}) + \frac{b_{m-1}}{b_0} \mathcal{K}_{m-1}(h_{1,1})  \right] \right\}\e^{\ii m\phi}.
\end{align}
To cast the equations above in a compact form, we have introduced the operators $\mathcal{K}_m$ and $\mathcal{L}_m$ as
\begin{equation*}\refstepcounter{equation}\label{eq:Lm}
\mathcal{K}_m (\cdot) = \frac{1}{r^2}\left\{\partial_r\left[ r (\cdot) \right] - 2m (\cdot)\right\}, \quad \mathcal{L}_m (\cdot) = \frac{1}{r}\partial_r \left[ r\partial_r(\cdot) \right]-\frac{m^2}{r^2}(\cdot),\eqno{(\text{\theequation a,b})}
\end{equation*}
with $(\cdot)$ denoting their operands. We then deduce the following linear partial differential equations by using Eq.~\eqref{eq:Laph1} in  Eq.~(\ref{Eq_h1PDE}a) and collecting $O(1)$ and $O(\delta)$ terms as follows: 
\begin{align}
\mathcal{F}(p_0)&= \left(1-2r^2\right)u_0-r u_1\e^{\ii\phi} + v'\left(q_0 + q_1\e^{\ii\phi}  - \frac{b_0(1-r^2)}{2v}\right),  \\
 \mathcal{F}(p_1)&+ 3\widetilde{\mathcal{F}}(p_0) +\sum_{m=2}^M\frac{b_m}{b_0}\left[h_{0,0}^3\left(\frac{m^2(3-2m) p_0 + \ii m^2\partial_\phi p_0}{r^2} + \frac{m^2\partial_r p_0 -2m\ii \,\partial_r\partial_\phi p_0}{r}\right)-\frac{m\ii\, \partial_\phi p_0\,\partial_r h_{0,0}^3}{r} \right]\e^{\ii m \phi}\nonumber\\
=&-\sum_{m=1}^M \left[ u_m r^m + \frac{(m-1)b_{m-1}u_1r}{2b_0} -\frac{(m+1)b_{m+1}(2r^m-r)u^*_1}{2b_0} - \frac{(m+3)r^m-2(m+1)r^2+m-1}{b_0}b_mu_0\right]\e^{\ii m\phi}\nonumber\\
&+v'\sum_{m=2}^M \left[q_m-b_m\frac{ 2r^m+m-1-r^2(1+m) }{2v}\right]\e^{\ii m\phi},
\end{align}
where
\begin{equation}
\mathcal{F}(\cdot)  = \frac{1}{r}\partial_r \left[ rh_{0,0}^3\partial_r (\cdot) \right] + \frac{h_{0,0}^3}{r^2}\partial_\phi^2 (\cdot),\quad \widetilde\mathcal{F}(\cdot) = \frac{1}{r}\partial_r \left[ r\tilde{h}_{0}h_{0,0}^2\partial_r (\cdot) \right] + \frac{h_{0,0}^2}{r^2}\partial_\phi \left[\tilde{h}_0 \partial_\phi(\cdot)\right]
\end{equation}
and $\tilde h_0=h_0/a\vartheta-h_{0,0}$ corresponds to the azimuthal components of $h_0$ scaled by $a\vartheta$, see Eq.~\eqref{3D_LeadingOrderOuter}. Collecting $\phi$-dependent terms in this set of equations allows us to obtain the required boundary value problems for each of  $h_{1,0}$, $h_{1,1}$, $\tilde{h}_{1,m}$, $m\ge1$. Specifically, for the axisymmetric and centroid corrections we get, respectively,
\begin{subequations}\label{eq:BVPs}
\begin{align}
&\mathcal{F}_0\left[\mathcal{L}_0(h_{1,0})\right] + \left(2r^2-1\right)u_0 - v'\left[q_0  - \frac{b_0(1-r^2)}{2v}\right]=0,\label{eq:h10} \\
&\mathcal{F}_1\left[\mathcal{L}_1(h_{1,1})\right] +r u_1 - v'q_1 = 0,\label{eq:h11}
\end{align}
whereas the corrections $\tilde{h}_{1,m}$ for $m\ge1$ satisfy 
\begin{align}
\mathcal{F}_m&\left[\mathcal{L}_m(\tilde{h}_{1,m})\right] + T_{1,m} + T_{2,m} + T_{3,m} + T_{4,m} + T_{5,m}=0, \label{eq:h1m}       
\end{align}
\end{subequations}
where
\begin{equation}
T_{1,m} =  \chi_{m} u_mr^m + \frac{(m-1)b_{m-1}ru_1}{2b_0} -\frac{(m+1)b_{m+1}(2r^m-r)u^*_1}{2b_0}
\end{equation}
are terms coming from the $\partial_th$ term of Eq.~\eqref{GovPDE} and do not vanish as $r\to1^-$,
\begin{equation}
T_{2,m} = - v'\left[\chi_{m} q_m -\frac{b_m}{v}\left( h_{0,0}+h_{0,m} \right)\right]
\end{equation}
captures the leading contributions from the mass flux terms,
\begin{equation}
T_{3,m}=\frac{b_m}{b_0}\left\{ 3\widetilde\mathcal{F}_{m,0}(L_0)+m^2\mathcal{F}_m[\mathcal{K}_{m}(h_{1,0})] + m^2\mathcal{A}_m(L_0) -\left[(m+3)r^m-2(m+1)r^2+m-1\right]u_0\right\}\label{eq:T3}
\end{equation}
corresponds to corrections to the azimuthal disturbances induced by axisymmetric spreading, and 
\begin{align}
T_{4,m}&=\frac{b_{m-1}}{2b_0}\left\{ 3\chi_m\widetilde{\mathcal F}_{m-1,m}(L_1) + \left( m^2-1 \right)\mathcal{F}_m[\mathcal{K}_{m-1}(h_{1,1})]+(m-1)\left[m\mathcal{A}_{m-1}(L_1)+\mathcal{B}_{m-1}(L_1)\right]\right\},\\
T_{5,m}&=\frac{b_{m+1}}{2b_0}\left\{ 3\widetilde{\mathcal{F}}_{m+1,-m}(L_1^*) +\left( m^2-1 \right)\mathcal{F}_m[\mathcal{K}_{m+1}(h_{1,1}^*)]+(m+1)\left[m\mathcal{A}_{m+1}(L_1^*)-\mathcal{B}_{m+1}(L_1^*)\right]\right\}
\end{align}
capture corrections due to the centroid motion. Here, we have used $\chi_m=1$ if $m>1$ and $\chi_1=0$, with
\begin{align*}\refstepcounter{equation}\label{eq:Fm}
\mathcal{F}_m(\cdot)&= \frac{1}{r}\partial_r\left[ rh_{0,0}^3\partial_r\left( \cdot \right) \right] -\frac{m^2 h_{0,0}^3}{r^2}\left( \cdot \right),\quad &\widetilde\mathcal{F}_{m,k}(\cdot)=&\frac{1}{r}\partial_r\left[ rh_{0,0}^2h_{0,m}\partial_r\left( \cdot \right) \right] -\frac{k h_{0,0}^2h_{0,m}}{r^2}\left( \cdot \right),\tag{\text{\theequation a,b}}\\
\mathcal{A}_m (\cdot)&= h_{0,0}^3 \left[ \mathcal{K}_m(\cdot) +\frac{2}{r^2} (\cdot)\right],&\mathcal{B}_m(\cdot)=&\frac{1}{r}\partial_r\left[h_{0,0}^3 (\cdot) \right] + \frac{(m-3)h_{0,0}^3(\cdot)}{r^2},\tag{\text{\theequation c,d}}
\end{align*}
and $L_k(r)=\mathcal{L}_k(h_{1,k})$.  What is of interest here is not the actual solutions to Eqs.~\eqref{eq:BVPs}, but their asymptotics as $r\to1^-$, which can be obtained by following the general methodology outlined in Appendix~\ref{sec:asympt}.

\section{Asymptotics of $\partial_\nu h$}\label{sec:asympt}

Each of the problems in Eqs.~\eqref{eq:BVPs} is a partial differential equation for some function $\psi(r,\tau)$ of the form 
\begin{equation}
\mathcal{F}_m\left[\mathcal{L}_m \psi\right] + A_m(r,\tau) = 0,\label{eqpm}
\end{equation}
where the operators $\mathcal{F}_m$ and $\mathcal{L}_m$ only involve functions and derivatives with respect to $r$, see Eqs.~(\ref{eq:Lm}b) and (\ref{eq:Fm}). For each of these problems, $A_m(1,\tau)\neq0$ so that $\mathcal{L}_m\psi \sim A_m(1,\tau)/(1-r)$ as $r\to1^-$ and
\begin{equation}
-\partial_r\psi\sim A_m(1,\tau)\ln(1-r)+B_m(\tau),\quad\text{as}\quad r\to1^-.\label{eq:psi}
\end{equation}
This two-term asymptotic expansion requires determining $B_m(\tau)$, which can be accomplished by following a similar approach as in Part I. Specifically, for each  $m\ge0$, we multiply Eq.~\eqref{eqpm} by $rf_m(r)$ for some function $f_m(r)$ to be determined,  and integrate over $r$ from 0 to $1-\varepsilon$, $0<\varepsilon\ll1$.  We assume that $f_m(r)$ has (at worst) a simple pole at $r=1$ and that $rf_m(r)<\infty$ as $r\to0$ so that the boundary terms arising from repeated integrations by parts vanish and we find
\begin{equation}\label{3D_intermm}
\int_0^{1-\varepsilon} r\mathcal{L}_m(\psi)\mathcal{F}_m(f_m)\,\mathrm{d}r = -\int_0^{1-\varepsilon} rA_m(r,\tau)f_m(r)\,\mathrm{d}r.
\end{equation}
Then, by making suitable choices for $f_m(r)$, we can isolate the required $B_m(\tau)$ terms needed for the asymptotics in Eq.~\eqref{eq:psi}, and, consequently, the asymptotics of Eqs.~\eqref{eq:BVPs}.
\subsection{The case when $m>0$}
When $m>0$,  if we require that 
\begin{equation}
\mathcal{F}_m(f_m)+r^{m}=0,\label{eqfm}
\end{equation}
we may perform additional integrations by parts in Eq.~\eqref{3D_intermm} to simplify its left-hand side so that, by substituting the asymptotics of $\partial_r\psi$, Eq.~\eqref{eq:psi}, we get
\begin{equation}
A_m(1,\tau)\ln\varepsilon + B_m(\tau) + O(\varepsilon\ln\varepsilon) =  -\int_0^{1-\varepsilon}r A_m(r,\tau)f_m(r)\,\mathrm{d}r.\label{prebetam}
\end{equation}
We then replace $\ln\varepsilon$ in Eq.~\eqref{prebetam} by
\begin{equation}
\ln\varepsilon = -\int_0^{1-\varepsilon}\frac{1}{1-r}\,\mathrm{d}r,
\end{equation}
make some term re-arrangement and take the limit $\varepsilon\to0$ in order to obtain $B_m(\tau)$ in integral form, namely
\begin{equation}
B_m(\tau)= \int_0^1 \left(\frac{A_m(1,\tau)}{1-r}-rA_m(r,\tau)f_m(r)\right)\,\mathrm{d}r.\label{eq:Bmt}
\end{equation}
Therefore, for given $f_m(r)$ satisfying Eq.~\eqref{eqfm} with $f_m(0)=0$ and $f_m\sim 1/(1-r)$ as $r\to1$,  $B_m(\tau)$  may be evaluated numerically for $m>0$. The solution to  Eq.~\eqref{eqfm} subject to these conditions is 
\begin{equation}
f_m(r)=\frac{4r^m}{(m+4)(1-r^2)^2}\left[\frac{g_m(r^2)}{g_m(1)}-1\right],\quad m\ge1
\end{equation}
where $g_m(r)$ is the Gauss hypergeometric function
\begin{equation}
 g_m(r) = {}_2F_1\left(\frac{m-1-\sqrt{m^2+9}}{2},\frac{m-1+\sqrt{m^2+9}}{2};m+1;r\right).
\end{equation}
Therefore, using the corresponding $A_m(r,\tau)$ for the two-term expansions of $\partial_r h_{1,1}$ and $\partial_r \tilde{h}_{1,m}$, we find
\begin{align}
-\partial_r h_{1,1} &\sim  u_1\ln(1-r) +u_1\beta_1 + v'\zeta_1\label{eq:h11as}\\
-\partial_r\tilde{h}_{1,m}&\sim\left( \chi_m u_m +\frac{m-1}{2b_0}b_{m-1}u_1-\frac{m+1}{2b_0}b_{m+1}u_1^*\right)\ln(1-r) +\chi_m\beta_mu_m + v'\zeta_m \nonumber\\
&+ \frac{m-1}{2b_0}\gamma_mb_{m-1}u_1-\frac{m+1}{2b_0}(2\beta_m-\gamma_m)b_{m+1}u_1^*  - \tilde{\beta}_m,\quad m\ge1,\label{eq:h1mas}
\end{align}
where we define
\begin{subequations}\label{eq:integrals}
\begin{align}
\beta_m &= \int_{0}^{1} \left( \frac{1}{1-r}-f_m(r)r^{m+1} \right)\,\mathrm{d}r,\label{eq:betam}\\
\gamma_m &=\int_0^1 \left(\frac{1}{1-r}-f_m(r)r^2\right)\,\mathrm{d}r,\label{eq:gammam}\\
\zeta_m &=\int_0^1 f_m(r)r\left\{q_m -\frac{b_m}{v}\left[ r^m-1+\frac{m+1}{2}\left( 1-r^2\right)\right]\right\}\,\mathrm{d}r,\label{eq:zetam}\\
\tilde\beta_m&=\int_{0}^{1} f_m(r)r\left( T_{3,m}+T_{4,m}+T_{5,m} \right) \,\mathrm{d}r.\label{eq:tbetam}
\end{align}
\end{subequations}
Formally, this methodology allows us to extract the asymptotics of Eqs.~(\ref{eq:BVPs}b) and (\ref{eq:BVPs}c). Although this does not pose major difficulties with Eq.~(\ref{eq:BVPs}b),  the integral term $\tilde{\beta}_m$ arising from the asymptotics of (\ref{eq:BVPs}c) is considerably involved and requires the full solutions for $h_{1,0}(r,\tau)$ and $h_{1,1}(r,\tau)$ everywhere in the domain $0\le r\le1$. In Ref.~\cite{Savva2018}, where we have $\zeta_m\equiv0$ (no mass flux), the $\tilde{\beta}_m$ were also omitted and only the $\beta_m$ and $\gamma_m$ were accounted for, which come from terms in Eqs.~(\ref{eq:BVPs}a) and (\ref{eq:BVPs}b) that do not vanish as $r\to1^-$. This omission did not affect the generally excellent agreement observed between the full numerical solution and the asymptotics. Although this point was not discussed in detail in Ref.~\cite{Savva2018}, insights about why doing so may be justified are offered at the end of Sec.~\ref{3DSec_MatchedAsymptotics}. 

\subsection{The case when $m=0$}

The case when $m=0$ follows in a similar manner, but we must also ensure that $\psi=h_{1,0}$ satisfies $\int_0^1rh_{1,0}\,\mathrm{d}r=0$, as required from the Fourier series expansion of Eq.~(\ref{Eq_h1PDE}b). Given the freedom we can afford in choosing what $\mathcal{F}_0(f_0)$ equals to in Eq.~\eqref{3D_intermm}, it turns out that by letting $\mathcal{F}_0(f_0)=1-2r^2$, we can explicitly enforce this integral condition in the subsequent integrations by parts applied to Eq.~\eqref{3D_intermm} to yield the corresponding Eq.~\eqref{prebetam}. In this case, solving $\mathcal{F}_0(f_0)=1-2r^2$ with the same conditions satisfied by $f_m(r)$, namely that $rf_0<\infty$ as  $r\to0$ and $f_0(r)$ has at worst a simple pole at $r=1$, gives
\begin{equation}
f_0(r)=\frac{2r^2}{1-r^2},
\end{equation}
so that we may deduce $B_0(\tau)$ from \eqref{3D_intermm} as
\begin{equation}
B_0(\tau)=\int_0^1 \frac{1}{1-r} \left(A_0(1,\tau) - \frac{2r^3 A_0(r,\tau)}{1+r}\right)\,\mathrm{d}r.
\end{equation}
Hence, using $A_0(r,\tau)= (2r^2-1)-v'[q_0-b_0(1-r^2)/(2v)]$ gives
\begin{equation}
-\partial_r h_{1,0} = u_0\ln(1-r)+\beta_0u_0 + v'\zeta_0\quad\text{as}\quad r\to1^-,\label{eq:h10as}
\end{equation}
where
\begin{equation}
\beta_0 = 2+\ln 2\quad\text{and}\quad \zeta_0 = -\frac{b_0}{4v}+ 2\int_0^1\frac{r^3q_0}{1-r^2}\,\mathrm{d}r.
\end{equation}

\section{Asymptotics of the integral term $B_m(\tau)$}\label{sec:integral}

Here we focus on the integrals for $B_m(\tau)$ when $\psi=\tilde{h}_{1,m}$, see Eqs.~\eqref{eq:integrals}, which are found to monotonically increase with $m$. These integrals cannot be computed analytically, but insights may be gained from their large-$m$ asymptotics.  Through a simple change of variable, Eqs.~\eqref{eq:betam} and \eqref{eq:gammam} may be equivalently written as
\begin{align}
\beta_m &= \ln 2 +\int_0^1 \left[\frac{1}{1-r} -\frac{2r^m}{(1-r)(m+4)}\left( \frac{g_m(r)}{g_m(1)}-1 \right) \right]\,\mathrm{d}r,\label{eq:betaint}\\
\gamma_m &=\ln 2 +\int_0^1 \left[\frac{1}{1-r} -\frac{2r^{(m+1)/2}}{(1-r)(m+4)}\left( \frac{g_m(r)}{g_m(1)}-1 \right) \right]\,\mathrm{d}r,\label{eq:gammaint}
\end{align}
whereas the integrals $\zeta_m$ and $\tilde{\beta}_m$, Eqs.~\eqref{eq:zetam} and \eqref{eq:tbetam}, respectively, may be cast in the form
\begin{equation}
I_m =\int_0^1 \frac{2r^{m/2}G(\sqrt{r})}{(1-r)(m+4)}\left( \frac{g_m(r)}{g_m(1)}-1 \right) \,\mathrm{d}r\label{eq:iint}
\end{equation}
for some appropriately given $G(r)$. Just as the integrands in Eqs.~\eqref{eq:zetam} and \eqref{eq:tbetam}, $G(r)$ is assumed to have a simple root at $r=1$. By taking the limit as $m\to\infty$, the values of these integrals are dominated by the contributions near $r=1$ due to the presence of $r^m$ and $r^{m/2}$ terms. Also important is the fact that  $g_m(r)$ also exhibits a boundary layer near $r=1$ as $m\to\infty$. To see this, consider the ordinary differential equation satisfied by $g_m(r)$ namely
\begin{equation}
\bar\varepsilon r(1-r)\partial_r^2 g_m + (1-r+\bar\varepsilon)\partial_rg_m + \left( \frac{1}{2}+2\bar\varepsilon \right)g_m=0,\label{eq:gm}
\end{equation}
where $\bar\varepsilon=1/m\ll1$. Introducing the rescaling 
\begin{equation}
\xi = \frac{1-r}{2\bar\varepsilon}=\frac{m}{2}(1-r)\label{eq:xi}
\end{equation}
allows us to probe into the inner region of Eq.~\eqref{eq:gm} whose leading-order solution, which is compatible with the outer solution,  $g_m\sim\sqrt{1-r}$, is
\begin{equation}
g_m=\hat{g}_m\xi\e^{\xi}\operatorname{K}_1(\xi),
\end{equation}
where $\operatorname{K}_1(\xi)$ is the modified Bessel function of the second kind and order unity. The constant $\hat{g}_m$ needs to be determined by matching, but in our case this constant is immaterial since the integrands contain divisions by $g_m(1)$ and we know that $\lim_{\xi\to0} \xi\e^\xi\operatorname{K}_1(\xi)=1$. 

Based on the above arguments, the dominant contributions to Eqs.~\eqref{eq:betaint}--\eqref{eq:iint} as $m\to\infty$ come from their integrands near $r=1$. Hence, we change the variable of integration to $\xi$ as given by Eq.~\eqref{eq:xi} and make the substitution  $g_m(\xi)/g_m(1)=\xi\e^{\xi}\operatorname{K}_1(\xi)$, so that we may deduce the large-$m$ asymptotics of Eqs.~\eqref{eq:betaint}--\eqref{eq:iint} as
\begin{align}
\beta_m &\sim \ln2 + \int_0^{m/2} \frac{1}{\xi}\left[ 1-\frac{\e^{-2\xi}}{\xi}\left( \xi\e^\xi\operatorname{K}_1(\xi)-1 \right) \right]\,\mathrm{d}\xi,\\
\gamma_m &\sim \ln2 + \int_0^{m/2} \frac{1}{\xi}\left[ 1-\frac{\e^{-\xi}}{\xi}\left( \xi\e^\xi\operatorname{K}_1(\xi)-1 \right) \right]\,\mathrm{d}\xi,\\
I_m &\sim \int_0^{m/2} \frac{\e^{-\xi} G\left(\sqrt{1-\frac{2}{m}\xi}\right) }{\xi^2}\left(\xi\e^\xi\operatorname{K}_1(\xi)-1\right)\,\mathrm{d}\xi,\label{eq:Im}
\end{align}
where we used the fact that $\lim_{m\to\infty}(1-\frac{2}{m}\xi)^{m/2}=\e^{-\xi}$. With the help of computer algebra software, it may be shown that
\begin{equation}
\beta_m\sim \ln m + 3\ln2-1+\gamma, \qquad \gamma_m \sim \ln m  + \frac{\pi}{2}-1+\gamma,\label{eq:est1}
\end{equation}
where $\gamma$ is the Euler--Mascheroni constant. The leading-term asymptotics of $I_m$  as $m\to\infty$ may be obtained  from the asymptotics of $G$ as $m\to\infty$ and computing the resulting integral with the upper limit of integration in Eq.~\eqref{eq:Im} set to infinity.

To determine the values of $\tilde{\beta}_m$ at each time step, we require the solution of $h_{1,0}(r,\tau)$ and $h_{1,1}(r,\tau)$, given by Eqs.~\eqref{eq:h10} and \eqref{eq:h11}. Since these depend on $v'$, the computational efficiency of the sought asymptotic model would have been severely compromised if we were to compute $h_{1,0}$ and $h_{1,1}$ at each time step. Instead, we have opted to use the $h_{1,0}$ and $h_{1,1}$ as computed with $v'\equiv0$, but incorporate into the final result of $\tilde{\beta}_m$ the large-$m$ asymptotics corresponding to $O(v'b_m)$ terms, which may be deduced as
\begin{align}
\tilde{\beta}_m \sim m&\left[3 \frac{v'}{v} \left(\zeta_0 b_m +\frac{\zeta_1 b_{m-1} + \zeta_1^* b_{m+1}}{2}\right)-\frac{u_0 b_m}{b_0}\left( 3\ln m +3\gamma+\pi -4\ln2 -8 +2\kappa \right)\right.\nonumber\\
&\left.-\frac{ u_1b_{m-1}+u_1^*b_{m+1}}{2}\left(  3\ln m+\pi-\ln2-3\beta_1+3\gamma+3\kappa-2 \right)\right]\label{eq:tildebeta_as}
\end{align}
with
\begin{equation}
 \kappa = \!\int_0^\infty\! \frac{\left(\xi\operatorname{K}_1(\xi) - \e^{-\xi}\right)\left(1-\e^{-\xi} \right)}{\xi^2}\,\mathrm{d}\xi\approx 0.5086.
\end{equation}
In other words, we write $\tilde{\beta}_m$, $m\ge1$, in the form 
\begin{equation}
\tilde{\beta}_m=m\tilde{\beta}^0_mu_0 \frac{b_m}{b_0} + (m+1)\tilde{\beta}^+_{m} u_1^*\frac{b_{m+1}}{2b_0}+(m-1)\tilde{\beta}^-_m u_1\frac{b_{m-1}}{2b_0}+3 \frac{v'}{v} \left[m\zeta_0 b_m +\frac{\zeta_1 (m-1) b_{m-1} + (m+1) \zeta_1^* b_{m+1}}{2}\right],\label{eq:tildebeta}
\end{equation}
where the last term only corresponds to the large-$m$ asymptotics of the $O(v')$ terms, whereas the parameters $\tilde{\beta}_m^0$ and $\tilde{\beta}_m^\pm$ are determined by setting $v'=0$ in each of the following 
\begin{equation}
\tilde{\beta}^0_m \frac{m b_m u_0}{b_0}= \int_0^1 rf_m T_{3,m}\,\mathrm{d}r,\quad \tilde{\beta}^-_m \frac{(m-1)u_1b_{m-1}}{2b_0}= \int_0^1 rf_m T_{4,m}\,\mathrm{d}r\quad\text{ and}\quad \tilde{\beta}^+_m \frac{(m+1)u^*_1b_{m+1}}{2b_0}= \int_0^1 rf_m T_{5,m}\,\mathrm{d}r. \label{eq:allbetas}
\end{equation}
For the last term in Eq.~\eqref{eq:tildebeta}, we have used a form which is asymptotically consistent with Eq.~\eqref{eq:tildebeta_as}, but better matches the structure of other terms in Eq.~\eqref{3D_ODE}. Although the inclusion of such terms may be arguably be perceived to be \emph{ad hoc}, their presence was not seen to appreciably impact the dynamics (see also Sec.~\ref{sec:outer}). Hence one may alternatively choose to discard them altogether so that mass flux contributions only come from the key $T_{2,m}$ term and the corresponding $v'\zeta_m$ term which results from it, see Eq.~\eqref{eq:h1mas}.

\begin{figure}
\includegraphics[scale=1]{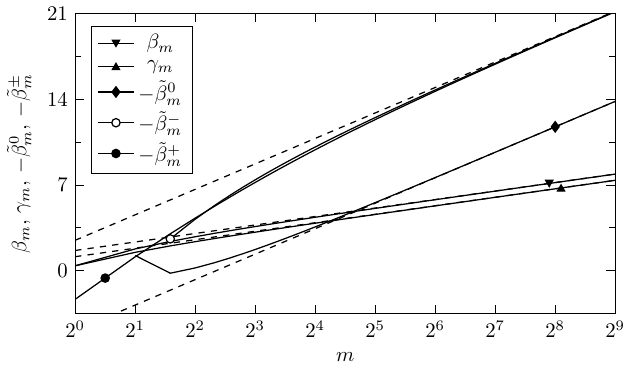}
\caption{The computed parameters for $\beta_m$, $\gamma_m$, $\tilde{\beta}_m^0$ and $\tilde{\beta}_m^\pm$ (solid curves).  Their large-$m$ asymptotics  are shown as dashed lines (see Eq.\eqref{eq:tildebeta_as}), namely $\beta_m\sim\ln m +3\ln2-1+\gamma$, $\gamma_m\sim \ln m +\pi/2-1+\gamma$, $-\beta_m^0\sim 3\ln m +3\gamma + \pi-4\ln2-8+2\kappa$ and 
$-\tilde{\beta}_m^\pm \sim 3\ln m+\pi-\ln2-3\beta_1+3\gamma+3\kappa-2$. The discrete values of these parameters are joined for clarity.}
\label{fig:betas}
\end{figure}

As noted above,  $T_{3,m}$, $T_{4,m}$ and $T_{5,m}$ in Eqs.~\eqref{eq:allbetas} are computed using the solutions $h_{1,0}$ and $h_{1,1}$ of Eqs.~\eqref{eq:h10}  and \eqref{eq:h11} when $v'\equiv0$, knowing also that $T_{3,1}=T_{4,1}=T_{4,2}=0$. In this manner, the way $\tilde{\beta}_m^0$ and $\tilde{\beta}_m^\pm$ are defined in Eqs.~\eqref{eq:allbetas} makes them only dependent on $m$ (when $v'=0$). Thus, their evaluation is carried out once using Gauss--Legendre quadrature, storing and retrieving their values whenever needed in simulations.  The integrand of $\tilde{\beta}^0_m$ requires knowledge of $h_{1,0}$, which can be written in terms of Spence's dilogarithm function, whereas those of $\tilde{\beta}^\pm_m$ require $h_{1,1}$, which is computed using a spectrally accurate method. A sufficiently large number of quadrature points was used to capture the logarithmic divergence of the integrands to compute the parameters to accuracy of at least four decimal places following the general principles detailed in Ref.~\cite{Savva2018}.  Similarly, for the time-dependent $\zeta_m(t)$, Eq.~\eqref{eq:zetam}, the same quadrature rule is used at each time step, where for efficiency in their evaluation we have pre-computed the $f_m(r)$ and re-used them at each time step. The result of the computation of $\beta_m$, $\gamma_m$, $\tilde{\beta}_m^0$ and $\tilde{\beta}_m^\pm$ is shown in Fig.~\ref{fig:betas} alongside with their asymptotics derived from Eqs.~\eqref{eq:est1} and \eqref{eq:tildebeta_as}.

\bibliography{Bibliography3D}{}
\end{document}